\documentclass[12pt]{article}
\usepackage[latin9]{inputenc}
\usepackage{amssymb,verbatim,amsmath}

\newtheorem{theorem}{Theorem}[section]

\newtheorem{remark}[theorem]{Remark}

\newtheorem{ex}{Example}[section]

\newtheorem{ass}{Assumption}[section]

\numberwithin{equation}{section}

\usepackage{color}

\begin{document}

\begin{center}
{\bf \Large One-dimensional flows of a polytropic gas: Lie group
classification, conservation laws, invariant and conservative
difference schemes}
\end{center}

\bigskip
\bigskip

%\date

%17.11.2020, RK

\bigskip

\begin{center}
{\large Vladimir A. Dorodnitsyn}$^a$,
{\large Roman Kozlov}$^b$
and
{\large Sergey V. Meleshko}$^c$
\end{center}

\bigskip

\noindent $^a$
Keldysh Institute of Applied Mathematics, Russian Academy of Science, \\
Miusskaya Pl.~4, Moscow, 125047, Russia; \\
{e-mail:  Dorodnitsyn@Keldysh.ru } \\
%  .... \\[10pt]
$^b$
Department of Business and Management Science, Norwegian School of Economics, \\
Helleveien 30, 5045, Bergen, Norway;  \\
{e-mail: Roman.Kozlov@nhh.no}  \\
%  .... \\[10pt]
$^c$
School of Mathematics, Institute of Science,
Suranaree University of Technology,
\\ 30000, Thailand; \\
{e-mail: sergey@math.sut.ac.th} \\
%  .... \\[10pt]

\bigskip
\begin{center}
{\bf Abstract}
\end{center}

\begin{quotation}
The paper considers one-dimensional flows of a polytropic gas in the
Lagrangian coordinates in three cases: plain one-dimensional
flows, radially symmetric flows  and spherically symmetric flows.
The one-dimensional flow of a polytropic gas is described
by one second-order partial differential equation in the Lagrangian variables.
Lie group classification of this PDE is performed. Its variational
structure allows to construct conservation laws with the help of
Noether's theorem. These conservation laws are also recalculated for
the gas dynamics variables in the Lagrangian  and Eulerian
coordinates. Additionally,  invariant and conservative difference
schemes are provided.
\end{quotation}

\bigskip
\bigskip

{\bf Key words:}

% Lagrangian gas dynamics

Polytropic gas

One-dimensional flows

% Radially/spherically symmetric flows

Lie point symmetries

Noether's theorem

Conservation law

Numerical scheme

\bigskip
\bigskip

%\eject

% {\bf \large Questions, notes, notations, what to do:}

% {\bf some notes  (2019)}

% Where  $ \gamma \protect \neq -1 $ comes from?

% {\bf SM:} "For $ \gamma = -1 $ the group is different.
% This case appeared in solving determining equations.
% This case is also studied, but I didn't include it.
% Is it necessary?
% Actually the case $ \gamma = -1 $ is used
% in cosmological theory.
% It is called Chaplygin gas."

% \bigskip

% Choice of signs  $\pm$  and coefficients in CL

% \bigskip

% $ n \neq 0 $ is valid for the whole paper

% \bigskip

% \begin{equation*}
% \gamma \neq  \gamma _* ={ n+3 \over n+1}
% \qquad
% \mbox{and}
% \qquad
% \gamma _* ={ n+3 \over n+1}
% \end{equation*}

% {\bf more recent (2020)}

% polytropic: how to describe?

% requirements for $\gamma \neq 0, -1, 1 ,  $

% define $\gamma _*$  once and not to repeat $ {n+3 \over n+1} $ everywhere

% potential $\varphi$: how to explain?

% references from [1]: are they for 1D or for any dimension?

% adiabatic: to explain?

% notations for additional symmetries
% and additional Noether symmetries

% Orthogonal mesh discussion

% \bigskip

% to cite Ibragimov (identity paper)

% recompute (for RK)

%\eject

\section{Introduction}

\label{Introduction}

Symmetries of the differential equations of mathematical physics are
their fundamental features. They reflect geometric structure of
solutions and physical principles of the considered models. We
recall that Lie group symmetries yield a number of  useful
properties   of differential equations (see
\cite{bk:Ovsiannikov1978, Ibragimov, bk:Olver[1986],
bk:MarsdenRatiu[1994], bk:Ibragimov[1999], bk:BlumanKumei[1989]}):

\begin{itemize}

\item

A  group action transforms the complete set of solutions into itself;
so it is possible to obtain new solutions from a given one;

\item

There exists a standard procedure to obtain
the whole set of invariants of a symmetry group of transformations;
it yields the forms of invariant solutions in
which they could be found (symmetry  reduction   of PDEs);

\item

For ODEs the known symmetry yields the reduction of the order;

\item

The invariance of  ODEs  and PDEs is
a necessary condition for the application of Noether's theorem
to variational problems to obtain conservation laws (first integrals for ODEs).

\end{itemize}

% If the differential equations have variational structure,
% i.e. they are Euler-Lagrange equations,
% then invariance of the Lagrangian function for a symmetry can be used to find
% conservation laws. This fundamental results is known as Noether's theorem.

% \bigskip
% \bigskip

The symmetry properties of the gas dynamics equations were studied
both in Eulerian coordinates \cite{bk:Ovsiannikov1978,
bk:Ovsiannikov[1994]} and in Lagrangian coordinates
\cite{bk:AkhatovGazizovIbragimov[1991], Ames, bk:HandbookLie_v2}.
Extensive group analysis of the one-dimensional gas dynamics
equations in mass Lagrangian coordinates was given
in~\cite{bk:AkhatovGazizovIbragimov[1991],  Ames,
bk:HandbookLie_v2}.
Here it should be also mentioned that
nonlocal conservation laws of
the one-dimensional gas dynamics equations in the mass Lagrangian
coordinates were found in~\cite{bk:SjobergMahomed2004}.
The authors of~\cite{bk:WebbZank[2009],bk:Webb2018} analyzed the Euler-Lagrange
equations corresponding to the one-dimensional gas dynamics
equations in the mass Lagrangian coordinates: extensions of the
known conservation laws were derived. These conservation laws
correspond to special forms of the entropy. The group nature of
these conservation laws is given in the present paper.

% Complete group
% analysis of the Euler-Lagrange equations of the one-dimensional gas
% dynamics equations of isentropic flows of a polytropic
% gas\footnote{The hyperbolic shallow water equations are equivalent
% to the gas dynamics equations for a polytropic gas with
% $\gamma=2$.} with $\gamma=2$ was given in~\cite{bk:SiriwatKaewmaneeMeleshko2016}.

As mentioned above, besides assisting with the construction of exact
solutions, the knowledge of an admitted Lie group allows one to derive
conservation laws. Conservation laws provide information on the basic
properties of solutions of differential equations.
They are also needed in the analyses of stability and global behavior of solutions.
Noether's theorem~\cite{bk:Noether[1918]} is the tool which relates
symmetries and conservation laws. However, an application of Noether's
theorem depends on the following condition: the differential equations
under consideration need to be presented  as Euler-Lagrange equations with
an appropriate Lagrangian,
i.e., Noether's theorem  requires variational structure.
There are also other approaches to find conservation laws,
which try to avoid this  requirement
\cite{bk:Shmyglevski, Ibragimov2[2007], bk:BlumanCheviakovAnco,
bk:SeligerWhitham[1968]}

Application of symmetries to difference and discrete equations
is more recent field of research
  \cite{Dorodnitsyn1989, Dorodnitsyn2011, Levi2006, Winternitz2011}.
One of its directions is discretization of differential equations
with preservation of Lie point symmetries. It is relevant to
construction of numerical schemes which inherit qualitative
properties of the underlying differential equations. This approach
was a base for a {series} of publications
 \cite{Dorodnitsyn1989, Dorodnitsyn1993b, BDK1997, DKW2000, DK2003, DKW2004},
which are summarized  in the book~\cite{Dorodnitsyn2011}. The method is
based on finding finite-difference invariants which  correspond to
the chosen mesh stencil and using them to construct invariant
difference equations and meshes. Recently this approach was applied
to shallow water  systems, wave equations and Green-Naghdi
system~\cite{DorKap1,  CDK, DorKapMel}.

The recent paper~\cite{DKM1} was devoted to Lie group
classification, conservation laws  and invariant difference schemes
of plain one-dimensional flows of a polytropic gas. Here we extend
these results to
 radially symmetric flows in two-dimensional space
and spherically symmetric flows in three-dimensional space. We refer
to all such flows as one-dimensional flows. The results
of~\cite{DKM1}  stand as a particular case in this paper.

There are two distinct ways to model phenomena in gas dynamics (see, e.g.
\cite{Samarskii,bk:RozhdYanenko[1978],Ovsiannikov}).
The typical approach uses  Eulerian coordinates, where
flow quantities (at each instant of time) are described
in fixed points.
%In this reference frame, one studies individual
%spatial positions, regardless of what particles reach those positions
%at a given instant of time.
Alternatively, the Lagrangian description
is used:   the particles are identified by the positions which
they occupy at some initial time.
In the Lagrangian description, there are also two ways to analyse the processes occurring in a gas.
{One of them uses a system of first-order PDEs for the gas dynamics variables.
The other approach uses a scalar second-order PDE to which this system can be reduced. }
%  In one of them the independent variables of the gas dynamic values are considered
% in Lagrangian coordinates, whereas in another way the gas dynamics equations
% are reduced to the equation for the function describing particle motion.
The latter way allows one to use  variational approach for analysis of the gas dynamics equations.
%One can consider the system of the gas dynamics
%equations, which are first-order partial differential equations
%\cite{Samarskii, bk:RozhdYanenko[1978], bk:DespresMazeran2005}.
%Alternatively, it is possible to reduce this system to one
%second-order partial differential equation \cite{bk:WebbZank[2009],
%bk:Webb2018}. This paper is devoted to Lie point symmetry analysis
%of this second-order PDE. We also use its variational structure to
%compute conservation laws.

The purpose of the paper is to present an overview of the authors'
results concerning the analysis of the gas dynamics equations of a
polytropic gas. It is devoted to symmetries, conservation laws and
construction of numerical schemes, which  preserve qualitative  properties of
the gas dynamics equations.

The article organized as follows. In the forthcoming  section we recall
Noether's theorem. Section~\ref{background} describers the gas dynamics equations,
their reduction to a single second-order PDE
and Lie point symmetries of this PDE. In Sections~\ref{arbitrary}
and~\ref{Special_cases} we consider the general case and the three
special cases of the Lie group classification. Invariance and
conservative properties of difference schemes are discussed in
Section~\ref{Difference_models}. Finally,  Section~\ref{Concluding}
presents concluding remarks.

%\eject

\section{Symmetries and Noether's theorem}

\label{Noether}

We briefly remind Noether's theorem~\cite{bk:Noether[1918]},
which will be used to find conservation laws
with the help of symmetries.
In the general case we have
several independent variables and dependent variables,
which are denoted as
$x=(x^{1},x^{2},\ldots,x^{n})$ and  $u=(u^{1},u^{2},\ldots,u^{m})$,
respectively.
All derivatives of order $k$ are denoted as $u_k$.

A point symmetry operator has the form
\begin{equation}    \label{operator_1}
X
=\xi^{i}\frac{\partial}{\partial x^{i}}
+ \eta^{k}\frac{\partial}{\partial u^{k}}
+\eta_{i}^{k}\frac{\partial}{\partial u_{i}^{k}}
+  \ldots  ,
\end{equation}
where we assume that  $\xi^{i}=\xi^{i}(x,u)$,
$ \eta^{k}=\eta^{k}(x,u)$ and that the operator is prolonged to
all derivatives $u_{i_{1} \ldots i_{l}}^{k}$ we need to consider.
We denote the  considered function  as   $  F ( x,u, u_1,  \ldots , u_k )  $.
It involves derivatives up to some finite order $k$.

% We denote all derivatives of some finite
% order as $p$ and present the considered equation as $ F ( x,u,p) = 0 $.

Noether's theorem is based on the identity~\cite{bk:Noether[1918],  Ibragimov}
\begin{equation}     \label{eq:Noether_1}
X F+FD_{i}\xi^{i} =  ( \eta^{k}-\xi^{i}u_{i}^{k} )  \frac{\delta
F}{\delta u^{k}} + D_{i}({N}^{i}F)  ,
\end{equation}
where
\begin{equation}   \label{eq:variational_1}
\frac{\delta}{\delta u^{k}}
= \sum_{s=0}  ^{\infty}
(-1)^{s}D_{i_{1}} \ldots D_{i_{s}}
\frac{\partial}{\partial u_{i_{1}i_{2} \ldots  i_{s}}^{k}},
\qquad
k=1,2, \ldots ,m,
\end{equation}
are variational operators,
and
\begin{equation}      \label{N_operator}
{N}^{i}
= \xi^{i}
+  \sum_{s=0}   ^{\infty}
D_{i_{1}} \ldots  D_{i_{s}}(  \eta^{k}-\xi^{i}u_{i}^{k}  )
\frac{\delta }{\delta u_{i i_{1} i_{2} \ldots  i_{s}}^{k}},
\qquad
i=1,2,  \ldots  ,n.
\end{equation}
The higher variational operators
${\displaystyle \frac{\delta }{\delta u_{i_{1} i_{2} \ldots  i_{s}}^{k}}}$
are obtained from the variational operators~(\ref{eq:variational_1})
by replacing $u^{k}$ with the corresponding
derivatives  $u_{i_{1} i_{2}  \ldots  i_{s}}^{k}$.

\begin{theorem}
(E.Noether) Let the Lagrangian function ${L}(x,u, u_1,  \ldots , u_k )$ satisfy  equation
%(\ref{eq:divergent})
\begin{equation}    \label{eq:divergent}
X{ L}+{L}D_{i}\xi^{i}=D_{i}B^{i}  .
\end{equation}
with any vector ${\bf B}=(B^{1},B^{2}, \ldots  ,B^{n})$ and a group
generator
\begin{equation*}
X   =  \xi^{i}(x,u)    \frac{\partial}{\partial x^{i}}
+\eta^{k}(x,u)    \frac{\partial}{\partial u^{k}},
\end{equation*}
then the generator $X$ is an admitted symmetry of the system of the
Euler-Lagrange equations %(\ref{eq:Euler-Lagrange-1}),
\begin{equation}      \label{eq:Euler-Lagrange-1}
\frac{\delta{L}}{\delta u^{k}}=0,
\qquad
k=1,2, \ldots ,m,
\end{equation}
and the vector
\begin{equation*}
({N}^{1}{L} - B^{1},
{N}^{2}{L}-B^{2},
\ldots ,
{N}^{n}{L}-B^{n})
\end{equation*}
is a conserved vector.
\end{theorem}

In the case ${\bf B}=(B^{1},B^{2}, \ldots ,B^{n})= {\bf 0} $, we call the symmetry $X$
variational symmetry,
otherwise we say that the symmetry $X$ is divergent.

It  is well-known  that variational and  divergent symmetries
are also symmetries of the Euler-Lagrange equations
 \cite{bk:Ovsiannikov1978, Ibragimov, bk:Olver[1986]}.
For Lie point symmetries,
i.e., symmetries with coefficients $\xi^{i}=\xi^{i}(x,u)$, $\eta^{k}=\eta^{k}(x,u)$,
and first-order Lagragians  $ L = L ( x, u, u _1   ) $
it  easily follows from the identities
\cite{bk:DorodnitsynKozlov[2011]}
% \footnote{For the
% sake of simplicity this identity is presented in the case, where a
% function $F$ does not depend on derivatives of order 2 and higher,
% and the coefficients do not depend on derivatives
% $\xi^{i}=\xi^{i}(x,u)$, $\eta^{k}=\eta^{k}(x,u)$. }:
%\footnote{In \cite{bk:DorodnitsynKozlov[2011]}, identity
%(\ref{eq:second})is given with $n=1$.}:
\begin{multline}
\frac{\delta}{\delta u^{j}}  \left(  X { L} + L D_{i}\xi^{i}  -D_{i}B^{i} \right)
=
X \left( \frac{\delta L}{\delta u^{j}} \right)
+\frac{\delta L}{\delta u^{k}}
\left(\frac{\partial\eta^{k}}{\partial u^{j}}
-\frac{\partial\xi^{i}}{\partial u^{j}}u_{i}^{k}
+\delta_{kj}D_{i}\xi^{i}\right),  \\
j=1,2,  \ldots ,m,
\label{eq:second}
\end{multline}
where $  \delta_{kj}$  is Kronecker symbol.

%\eject

\section{Equations of gas dynamics for one-dimensional flows}

\label{background}

We  consider three types of gas flows,
namely,  flows in  one-dimensional space,
radially symmetric flows   in two-dimensional space
and spherically  symmetric flows in three-dimensional space.
We will refer to these flows as {\it one-dimensional} flows.

The gas  is assumed to be  polytropic~\cite{Chernyi, Chorin, Ovsiannikov, Toro}.
For a polytropic gas the pressure $p$  and the density $ \rho $ are related as
\begin{equation}
p = S \rho  ^{\gamma} ,
\end{equation}
where variable $S$ is the function of the entropy $\tilde{S}$
\begin{equation*}         % \label{relation _2}
S=R  \   e^{(\tilde{S}-\tilde{S}_{0})/c_{v}}  .
\end{equation*}
Here $R$ is the gas constant,
$c_v$ is the specific heat capacity at constant volume
 and $\tilde{S}_{0}$ is constant.
The adiabatic constant is given as
 \begin{equation*}
\gamma = 1 + {  R  \over c_v} > 1  .
\end{equation*}
% is the ration of the specific heat capacity at constant pressure $c_p$
% to  the specific hear capacity at constant volume $c_v$.

We will also need the equation of state for the polytropic gas,
i.e.,    equation for the specific internal energy
\begin{equation}       \label{general_equation_of_state}
 \varepsilon  = \varepsilon  ( \rho ,  p) .
\end{equation}
It has the form
\begin{equation}       \label{equation_of_state}
\varepsilon    =   {  p   \over ( \gamma -1) \rho }  .
% \qquad
% \gamma   \neq 0, 1 .
\end{equation}
% For applications we can assume that $ \gamma > 1$.

The gas dynamics equations will be given in Eulerian and Lagrangian coordinates.
Eventually, they will be reduced to one scalar PDE of the second order,
which will be analyzed for admitted Lie point symmetries.

\subsection{Eulerian coordinates}

In  Eulerian coordinates $(t,r)$
the gas dynamics equation can be written as (see, e.g.
\cite{Ovsiannikov, Samarskii,bk:RozhdYanenko[1978]})
\begin{subequations}         \label{Euler_GD_system}
\begin{gather}
 \rho   _t   +   u   \rho _r   +  {  \rho   \over r^n  }    ( r^n u )  _r     =   0
 ,
% \qquad
% \left(
% \mbox{or}
% \qquad
% (r^{n}\rho)_{t}+(\rho r^{n}u)_{r}=0
%  \right)
  \label{Euler_GD_rho}
\\
 u   _t   +   u   u _r   +  {  1 \over  \rho   }    p_r     =   0  ,
% \qquad
% \left(
% \mbox{or}
% \qquad
% \rho(u_{t}+uu_{r})+p_{r}=0     \right)
  \label{Euler_GD_u}
\\
S_{t}+uS_{r}=0    .
  \label{Euler_GD_S}
\end{gather}
\end{subequations}
Here  we distinguish the case $n = 0$ with coordinate  $- \infty <r < \infty$  and
velocity $u$
from the cases $n = 1,  2 $  with radial distance from the origin $0< r < \infty$
and the radial velocity $u$.

We have $n = 0, 1, 2$
for  the plain one-dimensional flows,
the radially symmetric two-dimensional flows
and the spherically  symmetric three-dimensional flows, respectively.
Note that for these cases  $ n = d - 1$, where $d= 1, 2, 3$ is the space dimension.

We also use other representations of equation~(\ref{Euler_GD_S})
% can be replaced by the following equation
\begin{equation}      \label{Euler_GD_p}
p_t + u p_r + { \gamma p  \over  r^{n}  } (r^{n} u)_{r} = 0
\end{equation}
or
%by the equation
\begin{equation}      \label{Euler_GD_e}
\varepsilon  _t + u \varepsilon  _r + { p  \over  r^{n} \rho  } (r^{n} u)_{r} = 0  .
\end{equation}

% \begin{remark}
% The plane one-dimensional flows are included as the case  $n=0$.
% For them   $-\infty < r< \infty$ is the space coordinate
% and $u$ is the velocity.
% Three equations of plain one-dimensional gas dynamics (plain one dimensional flows)
% can be  presented  as three conservation laws, namely conservation laws for mass, momentum and energy.
% In the general case the equations~(\ref{Euler_GD_system})  have only conservation of mass and energy.
% For $n= 1$ and $n= 2 $ conservation of momentum
% is built into the differential equations which are obtained via symmetry reduction
% applied to the gas dynamics equations in the two- and three-dimensional space.
% \end{remark}

\subsection{Lagrangian  coordinates}

As well-known \cite{Samarskii, bk:RozhdYanenko[1978], Ovsiannikov},
the mass Lagrangian coordinate $s$ and the Eulerian coordinate $r$ of the particle $s$ are related by the formulas
\begin{equation}\label{eq:Nov16.1}
u=\varphi_t,
\qquad
 \rho=\frac{1}{\varphi^n\varphi_s},
\end{equation}
where $r=\varphi(t,s)$ defines the motion of a particle $s$.
Notice that for these $u $  and $\rho $
 equation~(\ref{Lagrange_GD_rho}) holds identically.

In the Eulerian coordinates $ (t,r) $ we can introduce the mass
Lagrangian coordinate $s$  as a  potential by the system
\begin{equation}\label{Change2}
 s_r =    r^{n} \rho ,
\qquad
 s_t  = -  r^{n} \rho u ,
\end{equation}
which is equivalent to  the 1-form
\begin{equation*}   % \label{Change_1}
 ds=  r^{n} \rho dr  -   r^{n} \rho u dt  .
\end{equation*}
Here we rely on the possibility to rewrite the equation~(\ref{Euler_GD_rho})
as  the conservation law
 \begin{equation*}
( r^{n}\rho)_{t}+(\rho r^{n}u)_{r}=0     ,
\end{equation*}
representing conservation of mass.
% ????This equation allows to introduce $ s $ as a potential.

% OR
% \begin{equation}\label{Change3}
%  s  = \int _{r_0} ^r      \rho (t,y)  y^n d y
% \end{equation}

In the mass Lagrangian  coordinates $(t,s)$,
which we will call Lagrangian  coordinates,
the time derivative  stands for the differentiation along  the {path}lines.
It is called the material derivative.
Total derivatives in the Lagrangian  coordinates $D^L _t $  and  $D  _s $
are related to those in  the Eulerian coordinates $D^E _t  $  and  $ D _r  $ as
\begin{equation}          \label{relation _3}
D^L _t = D^E _t   + u D _r   ,
\qquad
D  _s  =  { 1 \over   r^{n} \rho   } D _r    .
\end{equation}

%  {\bf (Is it derivative along  streamlines  ??? trajectories  ??? {path}lines ???). }

We rewrite gas the dynamics equations~(\ref{Euler_GD_system})
in the Lagrangian  coordinates $(t,s)$  as
\begin{subequations}    \label{Lagrange_GD_system}
\begin{gather}
\rho _ t  + \rho ^2 ( r^n u) _ s =  0   ,
% \qquad
% \left(
% \mbox{or}
% \qquad
%  \left( { 1 \over \rho } \right) _{t}  =  (  r^{n} u)_{s}     \right)
 \label{Lagrange_GD_rho}
\\
u_{t}   +  r^n p_{s}=0    ,
  \label{Lagrange_GD_u}
\\
S_{t}  =0  .
     \label{Lagrange_GD_S}
\end{gather}
\end{subequations}
We remark that here the gas dynamics variables
$\rho$, $u$,  $p$ and $S$ are functions
of the Lagrangian coordinates $t$ and $s$ while in the system
(\ref{Euler_GD_system}) they are functions
of the Eulerian coordinates $t$ and $r$.

The Eulerian spatial coordinate $r$  is a dependent variable in the Lagrangian
coordinates.
Equations (\ref{eq:Nov16.1}) can be rewritten in the form
\begin{equation}
r_t = u ,
\qquad
r_s = { 1 \over r^n \rho}   .
\end{equation}
It is also possible to use the 1-form
\begin{equation*}
 dr  =  {  ds \over r^{n} \rho } +    u dt  .
\end{equation*}

Notice that as for equation~(\ref{Euler_GD_S}), we also use other representations of
equation~(\ref{Lagrange_GD_S})
%We remark that the last equation~(\ref{Lagrange_GD_S}) can be replaced by the pressure
%equation
\begin{equation}       \label{Lagrange_GD_p}
p_t  +    \gamma  \rho p   (r^{n} u)_{s} = 0
\end{equation}
or
\begin{equation}      \label{Lagrange_GD_e}
\varepsilon  _t +  { p    } (r^{n} u)_{s} = 0  .
\end{equation}

\bigskip

%\subsection{The gas dynamics equation}

The Eulerian spatial coordinate
is  an additional dependent variable $   \varphi = r  $
in the Lagrangian coordinates $( t,s )$.
%Consider the equation of gas particle motion for the function $\varphi(t,s)$.
%This equation is derived from the gas dynamics equations by using the relations
%\begin{equation}      \label{potential}
%u = \varphi_{t}  ,
%\qquad
%\rho =  { 1 \over    \varphi ^n    \varphi_{s} }  .
%\end{equation}

%{\bf This formula already exists above!}
% For these $u $  and $\rho $
%given by~(\ref{potential})
% equation~(\ref{Lagrange_GD_rho}) holds identically.

% {\bf  To explain that such $ \varphi $ can be introduced !!!!)

% This is easy for $R^1$ where we have
% \begin{equation*}
% \left(  { 1 \over       \rho} \right) _ t = u _s
% \end{equation*}
% and introduce $ \varphi $  as a potential.

% In $R^n $ we have
% \begin{equation*}
% \left(  { 1 \over       \rho} \right) _ t = (  r^n u ) _s
% \end{equation*}
% and (equivalently)
% \begin{equation*}
% \left(  { 1 \over    r^n    \rho} \right) _ t = (  u ) _s
% \end{equation*}
% It is less clear how to introduce a potential.
% }

Equation~(\ref{Lagrange_GD_S}) can be solved
\begin{equation}            \label{S_solved}
S  = S(s),
\end{equation}
where $S(s)$ is an arbitrary function.

Using these results, it is possible to rewrite the last remaining
equation~(\ref{Lagrange_GD_u}),
as a partial differential equation of the second order
\begin{equation}     \label{GD_equation}
\varphi  _{tt}
+    \varphi^{ n  (1- \gamma)  }   \varphi_{s}^{ - \gamma }
\left(
S'   - n \gamma S  {   \varphi_{s} \over \varphi   }   - \gamma S   {   \varphi_{ss}
\over \varphi  _{s} }
\right)
= 0   .
\end{equation}
This PDE is called the {\it gas dynamics equation}  in the  Lagrangian
coordinates~\cite{Ovsiannikov, Chernyi}.

PDE~(\ref{GD_equation}) has a variational formulation,
namely,   it is  the Euler-Lagrange equation
\begin{equation}          \label{variational _short}
\frac{\delta{L}}{\delta\varphi}
= { \partial  L \over \partial \varphi }
-  D_t  ^L  \left( { \partial  L \over \partial \varphi_t }     \right)
-  D_s   \left( { \partial  L \over \partial \varphi_s }   \right)
 =    0
\end{equation}
for the Lagrangian
\begin{equation}           \label{Lagrange_function}
L    =   \frac{1}{2}\varphi_{t}^{2}
-\frac{S(s)}{\gamma-1}      \varphi^{(1-\gamma)n}  \varphi_{s}^{1-\gamma}    .
\end{equation}

\subsection{Conservation laws}        % and Noether's theorem}

We specify Noether's theorem,  given in Section~\ref{Noether},
 for PDE~(\ref{GD_equation}).
We consider Lie point symmetries of the form
\begin{equation}          \label{operator_2}
X =\xi^{t} (t,s,\varphi)
\frac{\partial}{\partial t}
+\xi^{s} (t,s,\varphi)
\frac{\partial}{\partial s}
+\eta^{\varphi} (t,s,\varphi)
\frac{\partial}{\partial\varphi}   .
\end{equation}
Such symmetries of the PDE~(\ref{GD_equation})
can be used to compute conservation laws
if they are also variational or divergence symmetries of the
Lagrangian~(\ref{Lagrange_function}).
We require that they satisfy
the condition of the
elementary action invariance \cite{Ibragimov}
\begin{equation}         \label{eq:Noether_2}
   X  {L}+{L}(D_{t} ^L \xi^{t}+D_{s}\xi^{s})
=D_{t}  ^L B _{1}+D_{s}  B _{2}
\end{equation}
for some functions $B_1 ( t, s, \varphi ) $ and $B_2 ( t, s, \varphi
) $. If  this condition  holds with $ B_1 = B_2 =0 $, then the
symmetry~(\ref{operator_2}) is called variational.
If~(\ref{eq:Noether_2}) holds with { trivial} (see~\cite{bk:Olver[1986]})
$ B_1 $ and $ B_2 $, then the symmetry  is divergent. We refer to
both variational and divergent symmetries as Noether symmetries.

Given a variational or divergent symmetry, we can find  the corresponding
conservation law
\begin{equation}    \label{CL_Lagrange}
D_{t} ^L ( T^{t} ) + D_{s} ( T^{s} )  =  0   ,
\end{equation}
where the densities are given by    the formulas
\begin{equation}       \label{conservation_laws_densities}
T^{t}
=\xi  ^{t} {L}
+  (  \eta^{\varphi} - \xi^{t}\varphi_{t} - \xi ^{s}\varphi_{s}  )  \frac{\partial
L }{\partial   \varphi_{t} }  - B_1   ,
\qquad
T^{s}
=\xi  ^{s}{L}
+  (  \eta^{\varphi} - \xi^{t}\varphi_{t} - \xi ^{s}\varphi_{s}  )  \frac{\partial
L }{\partial  \varphi_{s}}   -  B_2  .
\end{equation}

% %\eject

% \subsection{Conservation laws in Lagrangian  and Eulerian coordinates}

Conservation laws~(\ref{CL_Lagrange})
can be rewritten for the Eulerian coordinates as
\begin{equation}   \label{CL_Euler}
D_{t} ^E    ( ^{e} T^{t} ) + D_{r} (  ^{e} T^{r} ) =  0    .
\end{equation}
The relation
\begin{equation}    \label{conversion}
D_{t} ^L T^{t}+D_{s}T^{s}
=  \varphi_{s} \left(    D_{t}  ^E ( r^n \rho T^{t})    +    D_{r}(     r^n \rho u
T^{t}+T^{s} )    \right)
\end{equation}
can be proved by direct verification.
Therefore, if we have densities   $T^{t}$ and $ T^{s}$ of a conservation law in the
Lagrangian  coordinates,
we can find  the corresponding densities in the Eulerian coordinates as
\begin{equation}     \label{transformation_rule}
^{e}T^{t}= r^n \rho T^{t}  ,
\qquad
{}^{e}T^{r}= r^n \rho u T^{t} + T^{s}  .
\end{equation}

% {\bf Do we need the following ?:

% By the definition of velocity $u=\varphi_{t}$ and density
% $\rho=\frac{1}{\varphi^{n}\varphi_{s}}$,
% one has that
% \begin{equation}
% u_{r}= \varphi_{s}^{-1}    \varphi_{ts}
% \end{equation}
% }

\subsection{Equivalence transformations}

 PDE~(\ref{GD_equation}) contains an arbitrary function $S(s)$.
Thus, we need the group classification with respect to it.
The generators of the equivalence Lie  group has the form
\begin{equation}      \label{operator_3}
X^{e} = \xi^{t} \frac{\partial}{\partial t}
+\xi^{s} \frac{\partial}{\partial s}
+\eta^{\varphi} \frac{\partial}{\partial\varphi}
+\eta^{S} \frac{\partial}{\partial S}  .
\end{equation}

Computation gives { the generators of the equivalence  group. }
There   are five generators
\begin{multline}      \label{equivalence_general}
X_{1}^{e}=  {\partial  \over  \partial  t  } ,
\qquad
X_{2}^{e}=  {\partial  \over  \partial  s  } ,
\qquad
X_{3}^{e}=  t {\partial  \over  \partial  t  }   - 2  S {\partial  \over  \partial
S }    ,
\\
X_{4}^{e}=    s {\partial  \over  \partial  s   }   + ( 1 - \gamma  )   S {\partial
\over  \partial  S  }    ,
\qquad
X_{5}^{e}
=   \varphi  {\partial  \over  \partial   \varphi   }   + (  (n+1)   \gamma   - n
+1 )  S  {\partial  \over  \partial  S  }
\end{multline}
for the general case.
For $ n=0 $  there are two additional equivalence transformations
given by
\begin{equation}
X_{*, n }^{e}
=    {\partial  \over  \partial   \varphi   }
\qquad
\mbox{and}
\qquad
X_{**, n }^{e}
=  t   {\partial  \over  \partial   \varphi   }  .
\end{equation}
For the special values of the adiabatic exponent
$ \gamma _*  =   { n + 3 \over n+1} $
we obtain one additional generator
\begin{equation}      \label{equivalence_special}
X_{*, \gamma }^{e}
= t^2   {\partial  \over  \partial  t  }
+  t \varphi   {\partial  \over  \partial   \varphi  }  .
\end{equation}

\subsection{Group classification of the gas dynamics equation}

The Lie algebra of the admitted transformations is given by the generators
\begin{equation}      \label{symmetry_form}
X  =  \sum_{i=1}^{8}  k_{i}  Y _{i}  ,
\end{equation}
where
\begin{multline}
Y_{1}=  {\partial  \over  \partial  t  }    ,
\qquad
Y_{2}=   {\partial  \over  \partial  s }   ,
\qquad
Y_{3}=   {\partial  \over  \partial  \varphi  }   ,
\\
Y_{4}=  t {\partial  \over  \partial  t  }     ,
\qquad
Y_{5}=  s {\partial  \over  \partial  s }    ,
\qquad
Y_{6}=   \varphi   {\partial  \over  \partial   \varphi  }  ,
\\
Y_{7}=   t   {\partial  \over  \partial   \varphi  }  ,
\qquad
Y_{8}=     t^2    {\partial  \over  \partial  t   }  + t \varphi   {\partial  \over
\partial    \varphi  }    .
\end{multline}
The coefficients  $k_i$ satisfy the system
\begin{subequations}   \label{conditions}
\begin{gather}
(   k_{5}  s  +  k_{2}   )  S_{s}
=   (   -  2 k_{4}  +     (  1 - \gamma  )   k_{5}
+   (   ( n  + 1 )   \gamma   -  n  +  1    )  k_{6}    )   S  ,
\label{condition1}
 \\
 (  ( n+1)   \gamma  - n - 3 )   k_{8}   =  0  ,
\label{condition2}
 \\
n k_3   =  0  ,
\label{condition3}
 \\
n k_7   =  0  .
\label{condition4}
\end{gather}
\end{subequations}

For the general case  we get two admitted symmetries
\begin{multline}        \label{kernel1}
X_1   =   Y_1   =  {\partial  \over  \partial  t  }    ,
\qquad
X_2 =   {    (  ( n  + 1 )   \gamma   -  n  +  1  )    }    Y_4 + 2 Y_6
\\
=
 {     (  ( n  + 1 )   \gamma   -  n  +  1   )  }   t  {\partial  \over  \partial  t
 }
+  2  \varphi   {\partial  \over  \partial   \varphi  }      .
\end{multline}
For $ n= 0 $
there are two additional symmetries
\begin{equation}       \label{kernel3}
X_{*,  n}   = Y_{3} =   {\partial  \over  \partial   \varphi  }
\qquad
\mbox{and}
\qquad
X_{**,  n}   = Y_{7} =  t  {\partial  \over  \partial   \varphi  }      .
\end{equation}
For the special values  $ \gamma_*  = { n+3 \over n+1} $
there is one additional symmetry
\begin{equation}       \label{kernel2}
X_{* ,  \gamma }
= Y_{8}
=  t^2    {\partial  \over  \partial  t   }  + t \varphi   {\partial  \over
\partial   \varphi  }      .
\end{equation}

The condition~(\ref{condition1}) is the classifying equation for function  $S(s)$.
It can be rewritten as
\begin{equation}      \label{classification_equation}
( \alpha  s  + \beta)  S_s   = q S
\end{equation}
for some constants    $ \alpha $, $ \beta $  and $ q $.
This classifying equation was studied in~\cite{bk:AndrKapPukhRod[1998]}.
It was shown that one need to consider four cases of the entropy function  $S(s)$,
the general case and three special cases:

\begin{itemize}

\item

arbitrary $ S(s) $;

\item

$ S(s) = A_0 $, $A_0 = \mbox{const} $;

\item

$ S(s) = A_0 s^q $, $ q \neq 0$, $A_0 = \mbox{const} $;

\item

$ S(s) = A_0  e ^{ q s }$,  $ q\neq 0$, $A_0 = \mbox{const} $.

\end{itemize}

The same four cases were obtained for plain one-dimensional flows in~\cite{DKM1}.
Let us note that the equivalence transformations can be used to simplify these cases
to $ A_0 = 1$.

%\eject

\section{Arbitrary entropy $S(s) $}

\label{arbitrary}

Equation~(\ref{Lagrange_GD_rho}) can be rewritten in the form of
a conservation law as
\begin{equation*}
  \left( { 1 \over \rho } \right) _{t}  =  (  r^{n} u)_{s}     .
\end{equation*}
Thus, conservation of mass is included into the equations of the gas
dynamics system~(\ref{Lagrange_GD_system}).
Equation~(\ref{Lagrange_GD_S}) gives the conservation of the entropy
along {path}lines as the conservation law
\begin{equation*}
  S_t = 0     .
\end{equation*}

Let us examine the symmetries of the kernel of admitted
Lie algebras~(\ref{kernel1}),  (\ref{kernel3}) and~(\ref{kernel2})
for being variational or divergent  symmetries, which provide conservation laws.

% {\bf SM:}  Ovsiannikov told: do not call it "kernel", It should be "kernel of
% admitted Lie algebras"

\subsection{General case  $n \neq 0$,  $\gamma \neq { n+3 \over n+1} $  }

\label{arbitrary_general}

In the general cases the admitted symmetries~(\ref{kernel1})
provide us with one variational symmetry
\begin{equation}      \label{symmetry_general_variational}
Z_1 = X_{1}=   {\partial   \over  \partial  t }        .
\end{equation}
It leads to the conservation of energy with densities
\begin{equation}
T _1 ^t
=
  {  \varphi_{t}^{2}      \over 2 }
+  {S   \over  \gamma - 1 }     \varphi^{n(1-\gamma)}  \varphi_{s}^{1-\gamma}     ,
\qquad
T _1 ^s
=
   \varphi_{t}    S     \varphi^{n(1-\gamma)}  \varphi_{s}^{-\gamma}    .
\end{equation}

For the  gas dynamics variables this conservation law gets rewritten as
\begin{equation}
T _1 ^t
=
   {  u ^{2}  \over 2 }
 +  {S  \over  \gamma - 1 }     \rho  ^{\gamma - 1 }      ,
\qquad
T _1 ^s
=
     u     S  r^n  \rho ^{\gamma}     .
\end{equation}
In the Eulerian coordinates it  has the densities
\begin{equation*}
^e T _1 ^t
=
 r^n \left(   { \rho u ^{2}   \over 2 }
+  {S  \over  \gamma - 1 }     \rho  ^{\gamma}     \right)  ,
\qquad
^e T _1 ^r
=
r^n  \left(
  {  \rho u ^{2}    \over 2 }
+   {  \gamma  S  \over  \gamma - 1 }       \rho  ^{\gamma}     \right)   u   .
\end{equation*}

\subsection{Case $ n = 0 $, $\gamma \neq { n+3 \over n+1} $}

\label{n_special}

We get  one more variational symmetry
\begin{equation}     \label{Noether_symmetry_0_1}
Z_{*, n }  = X_{3}  =   { \partial \over \partial \varphi }  ,
\end{equation}
and  one divergent symmetry
\begin{equation}     \label{Noether_symmetry_0_2}
Z_{**, n} = X_{4}   =  t  { \partial \over \partial \varphi }
 \quad
 \mbox{with}   \quad  ( B _1  , B_2 ) =  ( \varphi  , 0 )     .
\end{equation}

These symmetries provide conservation laws
\begin{equation}
T_{*, n }^{t}= { \varphi_{t}^{2}  \over 2 }
+ { S   \over  \gamma-1  }    \varphi_{s}^{1-\gamma} ,
\qquad
T_{*, n }^{s}= S\varphi_{t}\varphi_{s}^{-\gamma}   ;
\end{equation}
\begin{equation}
T_{**, n }^{t}
=  \varphi  -  \varphi_{t}t  ,
\qquad
T_{**, n }^{s}= -  t S \varphi_{s}^{-\gamma}     ,
\end{equation}
representing  the conservation of momentum and the motion of the center of mass,
respectively.

In gas dynamics variables we can rewrite these conservation laws as
\begin{equation}   \label{conservation_energy}
T_{*, n }^{t}
=   { u^{2}   \over 2 }
+ { S   \over   \gamma-1  }   \rho^{\gamma-1}  ,
\qquad
T_{*, n }^{s}
=   S \rho^{\gamma} u   ;
\end{equation}
\begin{equation}       \label{conservation_center_of_mass}
T_{**, n }^{t}=  \varphi   -  t u   ,
\qquad
T_{**, n }^{s}=-    t   S \rho^{\gamma}   .
\end{equation}
Notice that the conserved vector $(T_{**, n }^{t}  ,T_{**, n }^{s})$ contains
the function $\varphi$.
In the Eulerian coordinates we get
\begin{equation*}
{}^{e}T_{*, n }^{t}
=    { \rho u^{2}  \over 2 }
+ {  S \rho^{\gamma}   \over  \gamma-1    }  ,
\qquad
{}^{e}T_{*, n }^{r}
= \left(  { \rho u^{2}  \over 2 }
+ {  \gamma S \rho^{\gamma}    \over  \gamma-1    }     \right) u  ;
\end{equation*}
\begin{equation*}
{}^{e}T_{**, n }^{t}=\rho( r - t u ) ,
\qquad
{}^{e}T_{**, n }^{r}=\rho u ( r - t u ) -  t S \rho^{\gamma}    .
\end{equation*}

\subsection{Special  case $n \neq 0$,  $ \gamma _* =  { n+3 \over n+1} $}

\label{arbitrary_special}

For $ \gamma =  \gamma _* $
the symmetries~(\ref{kernel1}) and~(\ref{kernel2})
lead to two variational symmetries:~(\ref{symmetry_general_variational}) and
\begin{equation}      \label{symmetry_special_variational}
% Z_1 = X_{1}=    {\partial   \over  \partial  t }    ,
\qquad
Z_{*,  \gamma }
=  {1 \over 2}  X_{2}=   2t   {\partial   \over  \partial  t }  +   \varphi
{\partial   \over  \partial  \varphi}
\end{equation}
and one divergence symmetry
\begin{equation}     \label{symmetry_special_divergence}
Z_{**,  \gamma  }   = X_{8}=   t^2  {\partial   \over  \partial  t }   +   t
\varphi  {\partial   \over  \partial  \varphi}
\qquad
\mbox{with}
\qquad
( B_1 , B_2 ) = \left( { \varphi ^2\over 2 } ,  0 \right)    .
\end{equation}
In addition to the conservation of energy, given in point~\ref{arbitrary_general},
there are conservation laws with densities
\begin{multline}
T _{*,  \gamma  }^t
=   2 t
\left(
 -  {  \varphi_{t}^{2}      \over 2 }
-  {S   \over  \gamma - 1 }     \varphi^{n(1-\gamma)}  \varphi_{s}^{1-\gamma}
 \right)
+ \varphi
  \varphi_{t}     ,
\\
T _{*,  \gamma   } ^s
=    -  2t
   S     \varphi^{n(1-\gamma)}   \varphi_{t}  \varphi_{s}^{-\gamma}
 +
   S     \varphi^{- n \gamma + n + 1 }  \varphi_{s}^{-\gamma}     ;
\end{multline}
\begin{multline}
T  _{**,  \gamma   } ^t
=   -  t^2
\left(
   {  \varphi_{t}^{2}      \over 2 }
+ {S   \over  \gamma - 1 }     \varphi^{n(1-\gamma)}  \varphi_{s}^{1-\gamma}
 \right)
 +  t  \varphi
  \varphi_{t}
- { \varphi ^2 \over 2 }   ,
\\
T  _{**,  \gamma  } ^s
=   -  t^2
    S     \varphi^{n(1-\gamma)}    \varphi_{t}   \varphi_{s}^{-\gamma}
 +  t
   S     \varphi^{ -  n \gamma + n + 1  }  \varphi_{s}^{-\gamma}      .
\end{multline}

We can rewrite these conservation laws for the gas dynamics variables
\begin{equation}
T _{*,  \gamma  } ^t
=  -  2t
\left(    {  u ^{2}  \over 2 }
+   {S  \over  \gamma - 1 }     \rho  ^{\gamma - 1 }
 \right)
+ r
   u      ,
\qquad
T  _{*,  \gamma  }^s
=    -  2t
   S  r^n  \rho ^{\gamma}  u
+
 S   r^{n+1}    \rho  ^{\gamma}      ;
\end{equation}
\begin{multline}
T  _{**,  \gamma  }^t
=    - t^2
\left(
  {  u ^{2}  \over 2 }
+   {S  \over  \gamma - 1 }     \rho  ^{\gamma - 1 }
 \right)
 +  t  r
   u
- { r ^2 \over 2 }   ,
\qquad
T _{**,  \gamma  }^s
=  -    t^2
    S  r^n  \rho ^{\gamma}  u
  +  t
 S   r^{n+1}    \rho  ^{\gamma}
\end{multline}
as well as in the Eulerian coordinates
\begin{multline*}
^e T _{*,  \gamma } ^t
=    -    2t
r^n \left(   { \rho u ^{2}   \over 2 }
+   {S  \over  \gamma - 1 }     \rho  ^{\gamma}     \right)
+
  r^{n+1}   \rho u       ,
\\
^e T  _{*,  \gamma  } ^r
=    -   2t
r^n  \left(
 {  \rho u ^{2}    \over 2 }
+   {  \gamma  S  \over  \gamma - 1 }       \rho  ^{\gamma}     \right)   u
  +
  r^{n+1}  (   \rho u ^{2}    +    S    \rho  ^{\gamma}    )     ;
\end{multline*}
\begin{multline*}
^e T  _{**,  \gamma  } ^t
=  - t^2
r^n \left(    { \rho u ^{2}   \over 2 }
+   {S  \over  \gamma - 1 }     \rho  ^{\gamma}     \right)
 +  t
   r^{n+1} \rho u
-   {r ^{n+2} \rho \over 2 }    ,
\\
^e T  _{**,  \gamma  } ^r
=  -    t^2
r^n  \left(
 {  \rho u ^{2}    \over 2 }
+   {  \gamma  S  \over  \gamma - 1 }       \rho  ^{\gamma}     \right)   u
  +  t
 r^{n+1}  (   \rho u ^{2}    +    S    \rho  ^{\gamma}    )
-  { r ^{n+2} \rho  u  \over 2 }    .
\end{multline*}

\subsection{Case $n =  0$, $ \gamma _* =  3 $}

\label{arbitrary_combined}

In this case the conservation law of the general case get extended
by  both the conservation laws given in point~\ref{n_special}
and by the conservation laws given in point~\ref{arbitrary_special}.

\section{Special cases of entropy}

\label{Special_cases}

Group classification of the PDE~(\ref{GD_equation}) gives three special cases
of the entropy function. They are examined in this section.
These cases  inherit the symmetries and   conservation laws  of the arbitrary  entropy
$S(s)$,
given in the preceding section.
We present only additional symmetries and conservation laws.

\subsection{Isentropic case $S(s) =A_0 $}

\label{constant}

In the Eulerian coordinates this case is  presented as
\begin{equation*}
S(r) =A_0
\qquad
\mbox{or}
\qquad
S_r = 0   .
\end{equation*}
For all cases
(the case of general $n$ and $\gamma$, the case $n=0$
and the case of special values $\gamma =\gamma_*$)
 there are two additional symmetries
% \begin{equation*}
% X_{1} = Y_1   =   {\partial   \over  \partial  t }    ,
% \qquad
% X_{2}  =  (  (n+1)  \gamma - n + 1 ) Y_3  + 2 Y_5
% \end{equation*}
% \begin{equation}      \label{symmetries_isentropic}
% =  (  (n+1)  \gamma - n + 1 )   t {\partial   \over  \partial  t }   + 2   \varphi
%{\partial   \over  \partial  \varphi}     ,
% \end{equation}
\begin{equation*}
X_{3}  = Y_2   =   {\partial   \over  \partial  s }   ,
\qquad
X_{4}  =   ( \gamma - 1 ) Y_4 - 2 Y_5 =
 ( \gamma - 1 )  t {\partial   \over  \partial  t }   - 2  s {\partial   \over
 \partial  s }     .
\end{equation*}

% For $\gamma _* = { n+3 \over n+1} $ there is one  additional symmetry
% \begin{equation}
% X_{*}  =  t ^2  {\partial   \over  \partial  t }
% +  t  \varphi  {\partial   \over  \partial  \varphi}     .
% \end{equation}

% In addition to the symmetries of the arbitrary S(s)
% (for all three both cases)
% there are two more symmetries,
% namely $X_3 $ and $X_4$.

\subsubsection{General  case $ n \neq 0$, $\gamma \neq  { n+3 \over n+1} $}

\label{constant_generic}

In the general case there are two additional variational symmetries
\begin{multline}
% Z_{1} = X_1   =   {\partial   \over  \partial  t }    ,
% \qquad
Z_{2} = X_3   =   {\partial   \over  \partial  s }   ,
\qquad
Z_{3}
=   {\gamma +1 \over 2 } X_2
+ { - ( n+1) \gamma + n + 3 \over 2 }   X_4
\\
=   (  (n+3) \gamma  - n - 1  )  t   {\partial   \over  \partial  t }
+    (  (n+1) \gamma  - n - 3  )  s   {\partial   \over  \partial  s }
+ ( \gamma +1 )   \varphi    {\partial   \over  \partial    \varphi   }   .
\end{multline}

The conservation laws of these case consist of the  conservation law
given in point~\ref{arbitrary_general}
(for arbitrary  $S(s)$) and the two additional ones, given by densities
\begin{equation}
T _2 ^t
=
  - \varphi_{s} \varphi_{t}     ,
\qquad
T _2 ^s
=
   { \varphi_{t}^{2}   \over 2 }
-  { \gamma S \over  \gamma - 1 }       \varphi^{n(1-\gamma)}
\varphi_{s}^{1-\gamma}    ;
\end{equation}
\begin{multline}
T _3 ^t
=  - (  (n+3) \gamma  - n - 1  )  t
\left(
  {  \varphi_{t}^{2}      \over 2 }
+  {S   \over  \gamma - 1 }     \varphi^{n(1-\gamma)}  \varphi_{s}^{1-\gamma}
 \right)
  -
(  (n+1) \gamma  - n - 3  )  s
 \varphi_{s} \varphi_{t}
\\
+
( \gamma +1 )   \varphi
  \varphi_{t}     ,
\\
T _3 ^s
=  -  (  (n+3) \gamma  - n - 1  )  t
    S     \varphi^{n(1-\gamma)}  \varphi_{t}  \varphi_{s}^{-\gamma}
 +
   (  (n+1) \gamma  - n - 3  )  s
\left(
   { \varphi_{t}^{2}   \over 2 }
-  { \gamma S \over  \gamma - 1 }       \varphi^{n(1-\gamma)}
\varphi_{s}^{1-\gamma}
 \right)
\\
+  ( \gamma +1 )
   S     \varphi^{ - n \gamma + n + 1 }  \varphi_{s}^{-\gamma}     .
\end{multline}

If rewritten for the gas dynamics variables, they take the form
\begin{equation}
T _2 ^t
=
  - { u \over r^n \rho }   ,
\qquad
T _ 2 ^s
=
   { u ^{2}   \over 2 }
-  { \gamma  S \over  \gamma - 1 }     \rho ^{\gamma -1}     ;
\end{equation}
\begin{multline}
T _3 ^t
= -   (  (n+3) \gamma  - n - 1  )  t
\left(
   {  u ^{2}  \over 2 }
+  {S  \over  \gamma - 1 }     \rho  ^{\gamma - 1 }
 \right)
 -
    (  (n+1) \gamma  - n - 3  )  s
 { u \over r^n \rho }
+
( \gamma +1 )    r
   u      ,
\\
T _3 ^s
=    -  (  (n+3) \gamma  - n - 1  )  t
     S  r^n  \rho ^{\gamma}  u
  +
   (  (n+1) \gamma  - n - 3  )  s
\left(
   { u ^{2}   \over 2 }
-  { \gamma  S \over  \gamma - 1 }     \rho ^{\gamma -1}
 \right)
\\
+ ( \gamma +1 )
  S   r^{n+1}   \rho  ^{\gamma}      .
\end{multline}

In the Eulerian coordinates these conservation laws have  densities
\begin{multline*}
^e T _ 2 ^t
=
  -  { u  }   ,
\qquad
^e T _ 2 ^r
=
-  { u ^{2}  \over 2 }
-  {  \gamma  S  \over  \gamma - 1 }      \rho  ^{\gamma-1 }        ;
\\
T _3 ^t
= -   (  (n+3) \gamma  - n - 1  )  t  r^n
\left(
   {   \rho u ^{2}  \over 2 }
+  {S  \over  \gamma - 1 }     \rho  ^{\gamma  }
 \right)
 -
    (  (n+1) \gamma  - n - 3  )  s
 { u  }
+
( \gamma +1 )    r^{n+1}  \rho    u      ,
\\
T _3 ^s
= =    -  (  (n+3) \gamma  - n - 1  )  t
     r^n    u \left(
   {   \rho u ^{2}  \over 2 }
+  {\gamma S  \over  \gamma - 1 }     \rho  ^{\gamma  }
 \right)
  -
    (  (n+1) \gamma  - n - 3  )  s
\left(
   { u ^{2}   \over 2 }
+  { \gamma  S \over  \gamma - 1 }     \rho ^{\gamma -1}
 \right)
\\
+ ( \gamma +1 )
 r^{n+1} ( \rho u^2 +  S \rho  ^{\gamma}  ) ,
\end{multline*}
 where $S(s)$  and $s$ is defined by system (\ref{Change2}).

% The components $ (T _3 ^t , T _3 ^s   )$ of the second additional conservation law
% cannot be presented in the Eulerian coordinates.

\subsubsection{Special cases}

For all special cases, namely
case  $ n =0  $,   $ \gamma \neq  { n+3 \over n+1} $
case    $ n \neq 0  $, $ \gamma_* =  { n+3 \over n+1} $
and case $ n =0  $,   $ \gamma_* = 3 $,
we get conservation laws of the arbitrary entropy $S(s)$,
which were  described in Section~\ref{arbitrary},
supplemented by the conservation law given in point~\ref{constant_generic}.

Note that
$$
Z_3 = Z_{*, \gamma } ,
$$
in other words for $\gamma = \gamma_* $ only
the second conservation laws from  point~\ref{constant_generic}  is new.

\subsection{Entropy case $S (s) =A_0 s^{q}$}

\label{power}

In the Eulerian coordinates
this entropy case is described by the differential constraint
\begin{equation}
S_{rr}=\rho^{-1}(nq\rho S+q\rho S_{r}r+q\rho_{r}Sr-\rho
S_{r}r)q^{-1}S^{-1}S_{r}r^{-1}   .
\end{equation}

% {\bf SM: }  I checked it

For all cases of arbitrary $ S(s) $ there is one additional symmetry
% The PDE~(\ref{GD_equation})  admits symmetries
% \begin{equation*}
% X_{1} = Y _1 =   {\partial   \over  \partial  t }    ,
% \end{equation*}
% \begin{equation}  \label{symmetries_power}
% X_{2}
% =   (   ( n+1)  \gamma -  n +1   ) Y_3   + 2 Y_5
% =    (   ( n+1)  \gamma -  n +1   )     {\partial   \over  \partial  t }
% + 2  \varphi     {\partial   \over  \partial  \varphi  }     ,
% \end{equation}
\begin{equation*}
 X_{3}
 = ( 1 - \gamma - q  ) Y_3   + 2 Y_4
 =   ( 1 - \gamma - q  )  t    {\partial   \over  \partial  t }    + 2   s
 {\partial   \over  \partial  s }    .
\end{equation*}

% For the special case $ \gamma _*  =  { n+3 \over n+1} $
% there is one additional symmetry
% \begin{equation}
% X_{*} = Y_6
% = t  ^2  {\partial   \over  \partial  t }
% +  t \varphi   {\partial   \over  \partial  \varphi}   .
% \end{equation}

\subsubsection{General case $ n \neq 0 $,  $ \gamma \neq { n+3 \over n+1} $}

\label{power_generic}

For the general case there is one  additional  variational symmetry
\begin{multline}
% Z_{1} = X _1 =   {\partial   \over  \partial  t }  ,
% \qquad
Z_2
=    {  ( n+1)  \gamma  - n - 3 \over 2 }     X _2  + {  \gamma + q + 1 \over 2 }
X_3
\\
= (  ( n+3 ) \gamma +2q - n - 1)  t  {\partial   \over  \partial  t }
+ ( ( n+1) \gamma - n - 3 )  s  {\partial   \over  \partial  s }
+  ( \gamma + q + 1 )   \varphi   {\partial   \over  \partial  \varphi}    .
\end{multline}
Thus, in addition to the conservation of energy given in~\ref{arbitrary_general}
we obtain the conservation law
\begin{multline}
T _2 ^t
=  -   (  ( n+3 ) \gamma +2q - n - 1)  t
\left(
   {  \varphi_{t}^{2}      \over 2 }
+  {S   \over  \gamma - 1 }     \varphi^{n(1-\gamma)}  \varphi_{s}^{1-\gamma}
 \right)
\\
  -  ( ( n+1) \gamma - n - 3 )  s
  \varphi_{s} \varphi_{t}
+
 ( \gamma + q + 1 )   \varphi
  \varphi_{t}     ,
\\
T _2 ^s
=    -  (  ( n+3 ) \gamma +2q - n - 1)  t
      S     \varphi^{n(1-\gamma)}   \varphi_{t}    \varphi_{s}^{-\gamma}
  +
  ( ( n+1) \gamma - n - 3 )  s
\left(
   { \varphi_{t}^{2}   \over 2 }
-  { \gamma S \over  \gamma - 1 }       \varphi^{n(1-\gamma)}
\varphi_{s}^{1-\gamma}
 \right)
\\
+  ( \gamma + q + 1 )
   S     \varphi^{ - n \gamma  + n + 1 }  \varphi_{s}^{-\gamma}     .
\end{multline}

For the gas dynamics variables it takes the form
\begin{multline}
T _2  ^t
=   -   (  ( n+3 ) \gamma +2q - n - 1)  t
\left(
   {  u ^{2}  \over 2 }
+  {S  \over  \gamma - 1 }     \rho  ^{\gamma - 1 }
 \right)
  -
   ( ( n+1) \gamma - n - 3 )  s
   { u \over r^n \rho }
\\
+
 ( \gamma + q + 1 )   r
    u      ,
\\
T _2 ^s
=     -   (  ( n+3 ) \gamma +2q - n - 1)  t
   S  r^n  \rho ^{\gamma}    u
  +
  ( ( n+1) \gamma - n - 3 )  s
\left(
   { u ^{2}   \over 2 }
-  { \gamma  S \over  \gamma - 1 }     \rho ^{\gamma -1}
 \right)
\\
+
 ( \gamma + q + 1 )
 S   r^{n+1}   \rho  ^{\gamma}       .
\end{multline}
To rewrite this conservation laws in the Eulerian coordinates
we use the relation
\begin{equation}      \label{s _for_power}
s = q r^n \rho { S \over S_r }
\end{equation}
to present the Lagrangian coordinate $s$.
This relation allows to write down the densities of the conservation law as follows
\begin{multline*}
^e T _2  ^t
=    -  (  ( n+3 ) \gamma +2q - n - 1)  t
r^n \left(    { \rho u ^{2}   \over 2 }
+ {S  \over  \gamma - 1 }     \rho  ^{\gamma}     \right)
  -
 ( ( n+1) \gamma - n - 3 )  {  q r^n \rho { S \over S_r }    }
 { u  }
\\
+
 ( \gamma + q + 1 )
   r^{n+1}  \rho u      ,
\\
^e T _2 ^r
=   -   (  ( n+3 ) \gamma +2q - n - 1)  t
r^n  \left(
  {  \rho u ^{2}    \over 2 }
+  {  \gamma  S  \over  \gamma - 1 }       \rho  ^{\gamma}     \right)   u
 \\
  -    ( ( n+1) \gamma - n - 3 )  {   q r^n \rho { S \over S_r }  }
\left(
  { u ^{2}  \over 2 }
+ {  \gamma  S  \over  \gamma - 1 }      \rho  ^{\gamma-1 }
 \right)
+
 ( \gamma + q + 1 )
 r^{n+1} (   \rho u ^{2}    +    S    \rho  ^{\gamma}    )     .
\end{multline*}

\subsubsection{Special case $ n = 0 $, $\gamma \neq {n+3 \over n+1}$}

\label{power-special_2}

For $ n =0  $  the additional Noether symmetries are the same as in the general
case.
Therefore, we get conservation laws
given in points~\ref{arbitrary_general}, \ref{n_special} and~\ref{power_generic}.

\subsubsection{Special case  $n \neq 0$,  $ \gamma _* = { n+3 \over n+1} $}

\label{power-special}

The special case of $ \gamma _*$ splits for  values of $q$.
For general $q$ we get the same Noether symmetries
% \begin{equation}
% Z_1  ,
% \qquad
% Z_* ,
% \qquad
% Z_{**}
% \end{equation}
as in the case of arbitrary  $S(s)$.
Therefore, we obtain the same conservation laws
as  given in points~\ref{arbitrary_general}   and~\ref{arbitrary_special}.

For the particular case
$q_* = - 2 \frac{n+2}{n+1}$
there is one additional variational symmetry
\begin{equation}    \label{symmetry_power_additional}
Z_{*, q  }
 = { 1 \over 2 } X_2
=    t    {\partial   \over  \partial  t }    +   s    {\partial   \over  \partial
s }    .
\end{equation}
It provides with the following conservation law
\begin{multline}
T ^t _{*, q  }
=    - t
\left(
  {  \varphi_{t}^{2}      \over 2 }
+  {S   \over  \gamma - 1 }     \varphi^{n(1-\gamma)}  \varphi_{s}^{1-\gamma}
 \right)
  -   s
 \varphi_{s} \varphi_{t}    ,
\\
T ^s   _{*, q  }
=   -   t
    S     \varphi^{n(1-\gamma)}  \varphi_{t}  \varphi_{s}^{-\gamma}
  +  s
\left(
   { \varphi_{t}^{2}   \over 2 }
-  { \gamma S \over  \gamma - 1 }       \varphi^{n(1-\gamma)}
\varphi_{s}^{1-\gamma}
 \right)  .
\end{multline}

It is also possible to present this conservation laws for the gas dynamics variables
\begin{multline}
T ^t  _{*, q }
=   -  t
\left(
   {  u ^{2}  \over 2 }
+   {S  \over  \gamma - 1 }     \rho  ^{\gamma - 1 }
 \right)
  -   s
 { u \over r^n \rho }  ,
\\
T ^s  _{*, q  }
=    -   t
   S  r^n  \rho ^{\gamma}     u
  +  s
\left(
   { u ^{2}   \over 2 }
-  { \gamma  S \over  \gamma - 1 }     \rho ^{\gamma -1}
 \right)    .
\end{multline}
To rewrite these densities in the Eulerian coordinates
we employ the relation~(\ref{s _for_power})
and obtain densities
% \begin{equation}
% s = q r^n \rho { S \over S_r }
% \end{equation}
% for the following
\begin{multline*}
^e T ^t  _{*, q  }
=  -   t
r^n \left(   { \rho u ^{2}   \over 2 }
+  {S  \over  \gamma - 1 }     \rho  ^{\gamma}     \right)
  -  {q r^n \rho { S \over S_r }  }
  { u  }   ,
\\
^e T ^r   _{*, q  }
=  -     t
r^n  \left(
  {  \rho u ^{2}    \over 2 }
+  {  \gamma  S  \over  \gamma - 1 }       \rho  ^{\gamma}     \right)   u
   -   { q r^n \rho { S \over S_r }  }
\left(
 { u ^{2}  \over 2 }
+ {  \gamma  S  \over  \gamma - 1 }      \rho  ^{\gamma-1 }
 \right)     .
\end{multline*}

\subsubsection{Case $n=0$, $ \gamma _* = 3$}

\label{power-combined}

We get the same conservation laws as described in the previous point.
Note that $ n= 0$ leads to  $ q_* = -4$.

%%%%%%%%%%%%%%%%%================================

\subsection{Entropy case $S (s) =A_0 e^{q s}$}

\label{exponential}

Let us note that this special case can be given in the Eulerian  coordinates
by the differential constraint
\begin{equation}        \label{s _for_exponent}
S_{r}=  q r^{n} \rho S  .
\end{equation}

For all cases of Section~\ref{arbitrary}
there is one additional symmetry
% There are three symmetries
\begin{equation}    \label{symmetry_exponential}
% X_{1} = Y_1   =   {\partial   \over  \partial  t }  ,
% \qquad
% X_{2}=    ( ( n+1 ) \gamma - n + 1 ) Y_3 + 2 Y_5
% \\
% =  ( ( n+1 ) \gamma - n + 1 )  t {\partial   \over  \partial  t }
% + 2   \varphi {\partial   \over  \partial  \varphi}   ,
% \qquad
X_{3}  = - 2 Y_2 + q Y_4
= q t {\partial   \over  \partial  t }   - 2   {\partial   \over  \partial  s}   .
\end{equation}

% values $\gamma$  and one additional symmetry
% \begin{equation}
% X_{*}=t^2     {\partial   \over  \partial  t }  + t \varphi    {\partial   \over
%\partial  \varphi}
% \end{equation}
%  for $ \gamma _* =  { n+3 \over n+1} $.

\subsubsection{General case $ n \neq 0 $, $\gamma \neq    { n+3 \over n+1} $}

\label{exponential-generic}

For the general case there is one additional variational symmetry
\begin{equation}
% Z_{1} = X_1   =   {\partial   \over  \partial  t }   ,
% \qquad
Z_{2}
=   { - ( n+1) \gamma + n + 3 \over 2}    X_2
+ {  q \over 2 }  X_3
 =  2q t  {\partial   \over  \partial  t }
+  (  ( n+1) \gamma -  n -  3  )       {\partial   \over  \partial  s }
+  q \varphi {\partial   \over  \partial  \varphi  }     .
\end{equation}
%Conservation law for   $ Z_{1}$ is given in~\ref{arbitrary_general}.
The supplementary conservation law has densities
\begin{multline}
T _2 ^t
=   -  2q t
\left(
 {  \varphi_{t}^{2}      \over 2 }
+  {S   \over  \gamma - 1 }     \varphi^{n(1-\gamma)}  \varphi_{s}^{1-\gamma}
 \right)
  -    (  ( n+1) \gamma -  n -  3  )
   \varphi_{s} \varphi_{t}
+
  q \varphi
  \varphi_{t}     ,
\\
T _2 ^s
=   -   2q t
     S     \varphi^{n(1-\gamma)}  \varphi_{t}  \varphi_{s}^{-\gamma}
  +
   (  ( n+1) \gamma -  n -  3  )
\left(
   { \varphi_{t}^{2}   \over 2 }
-  { \gamma S \over  \gamma - 1 }       \varphi^{n(1-\gamma)}
\varphi_{s}^{1-\gamma}
 \right)
\\
+
  q
  S     \varphi^{ - n \gamma + n + 1 }  \varphi_{s}^{-\gamma}     .
\end{multline}

For the  gas dynamics variables we get
\begin{multline}
T _2 ^t
=  -    2q t
\left(
   {  u ^{2}  \over 2 }
+  {S  \over  \gamma - 1 }     \rho  ^{\gamma - 1 }
 \right)
 -
  (  ( n+1) \gamma -  n -  3  )
{ u \over r^n \rho }
+
q r
 u   ,
\\
T _2 ^s
=    -   2q t
   S  r^n  \rho ^{\gamma}  u
  +
  (  ( n+1) \gamma -  n -  3  )
\left(
   { u ^{2}   \over 2 }
-  { \gamma  S \over  \gamma - 1 }     \rho ^{\gamma -1}
 \right)
+
  q
  S   r^{n+1}    \rho  ^{\gamma}      .
\end{multline}
Finally, we rewrite these densities in the Eulerian coordinates
\begin{multline*}
^e T _2 ^t
=    - 2q t
r^n \left(   { \rho u ^{2}   \over 2 }
+  {S  \over  \gamma - 1 }     \rho  ^{\gamma}     \right)
  -
  (  ( n+1) \gamma -  n -  3  )
  { u  }
+
q
  r^{n+1} \rho u   ,
\\
^e T _2 ^r
=     -   2q t
r^n  \left(
  {  \rho u ^{2}    \over 2 }
+  {  \gamma  S  \over  \gamma - 1 }       \rho  ^{\gamma}     \right)   u
  -
  (  ( n+1) \gamma -  n -  3  )
\left(
 { u ^{2}  \over 2 }
+ {  \gamma  S  \over  \gamma - 1 }      \rho  ^{\gamma-1 }
 \right)
\\
+    q
   r^{n+1} (   \rho u ^{2}    +    S    \rho  ^{\gamma}    )     .
\end{multline*}

\subsubsection{Special cases}

For all special cases we get the same additional conservation law as
in the general case of $n$ and $\gamma$.
We remark that because of
$$
Z_2 = q Z_{* , \gamma }
$$
the corresponding conservation law, given in point~\ref{exponential-generic},
is not new for the special values $ \gamma = \gamma_*$.

% \subsubsection{Special case $ n = 0 $}

% \label{exponential-special_2}

% For $ n =0  $  the additional Noether symmetries are the same
% as in the general case.
% Therefore, we get conservation laws
% given in points~\ref{arbitrary_general},\ref{n_special}
% and~\ref{exponential-generic}.

% \subsubsection{Special case $ \gamma _* =  { n+3 \over n+1} $}

% \label{exponential-special}

% The special case  $ \gamma _* $ leads to the same Noether symmetries
% \begin{equation}
% Z_1 ,
% \qquad
% Z_{* , \gamma = \gamma_*}  ,
% \qquad
% Z_{**, \gamma = \gamma_*}
% \end{equation}
% as in the case of arbitrary $S(s)$ and  $\gamma_*$
% because $ Z_2 = q Z_{* , \gamma = \gamma_*}$.
% Therefore, there are the same conservation laws
% as given in points~\ref{arbitrary_general} and~\ref{arbitrary_special}.

\subsection{Discussion}

Complete Lie group classification of the gas dynamics equation
in  the Lagrangian coordinates~(\ref{GD_equation})
allows us to find all  conservation laws which can be found using Noether's theorem
and admitted symmetries.
The group classification has three cases of the entropy
for which there exist additional symmetries.
In the Eulerian coordinates these three cases are defined by  differential
constraints of first or second order.
Notice that the overdetermined systems which consist of the gas dynamics equations
and one of the considered differential constraints are involutive.
The authors of~\cite{bk:WebbZank[2009],bk:Webb2018} also found conservation laws
corresponding to special forms of the entropy.
% {\bf is it for 1-D or for N-Dimensions ?}
Here the symmetry nature of these conservation laws is explained.

In contrast to~\cite{bk:SjobergMahomed2004}
the conservation laws, obtained in this paper,  are local. It should be also noted
that these
conservation laws are naturally derived:
their counterparts in Lagrangian coordinates were derived directly
using Noether's theorem without any additional assumptions.
In contrast to the two-dimensional Lagrangian gas dynamics
the special cases of the entropy in the Lagrangian coordinates are given explicitly.
In the two-dimensional case~\cite{Kaptsov_press} the entropy is arbitrary,
but the admitted symmetry operators  contain  functions
satisfying quasilinear partial differential equations.

%\eject

%\subsection{???Remarks on the conservation laws of the 1D gas dynamics equations}
%\begin{remark}

In a conservative form the one-dimensional gas dynamics equations
(\ref{Euler_GD_system})
are \cite{bk:RozhdYanenko[1978]}
\begin{subequations}
\begin{gather}
\left[  r^{n}\rho  \right]_t
+
\left[ r^{n}\rho u \right]_r
=0,
\label{eq:mass_original}
\\
\left[ r^{n}\rho u  \right]_t
+
\left[ r^{n}( \rho u^{2}  + p )\right]_r
=
nr^{n-1}p,
\label{eq:mom_original}
\\
\left[ r^{n}
\left(  \rho \varepsilon + \frac{\rho u^{2}}{2}\right)  \right]_t
+
\left[ r^{n}
\left( \rho \varepsilon+\frac{\rho u^{2}}{2} +  {p}   \right)  u
\right]_r
=0 ,
\label{eq:energy_original}
\end {gather}
\end{subequations}
where $ [ \ldots ] _t $ and  $ [ \ldots ] _t $ denote total derivatives with respect to time $t$
and the Eulerian coordinate $r$.
One notes that the equation corresponding to the conservation law of momentum
is not homogeneous. However, most methods for constructing conservation
laws can only construct homogeneous conservation laws.

{Consider}
%Here we analyze
inhomogeneous conservation laws of the one-dimensional
gas dynamics equations
\begin{equation}\label{eq:nov10.1}
D_{t} \left[ f^{t}  \right]  +D_{r}  \left[ f^{r}  \right] =f,
\end{equation}
where $D_{t}$ and $D_{r}$ are the total derivatives and the functions
$f^{t}$, $f^{r}$ and $f$ depend on $(t,r,\rho,u,p)$.
The method which is used to derive such conservation laws
consists of obtaining an overdetermined system of partial differential
equations for the functions $f^{t}$, $f^{r}$ and $f$ and finding
its general solution. The overdetermined system is derived by substituting
the main derivatives $\rho_{t}$, $u_{t}$, and $p_{t}$ found from
the gas dynamics equations into (\ref{eq:nov10.1}), and splitting it with respect to
the parametric derivatives.

Calculations show that the general solution of this system provides the
conservation laws
\begin{equation}
\left[  \rho  F  \right]_t
+
\left[ \rho  u F   \right]_r
=
\rho  \left(
F_{t} + u  \left( F_{r} - { n  \over r}  F \right)
\right),
\label{eq:mass}
\end{equation}
\begin{equation}
\left[  h \rho u  \right] _t
+
\left[ h(\rho u^{2}+p) \right]_r
=
h_{t}\rho u+h_{r}(\rho u^{2}+p)   -   { n \over r }  h \rho u^{2},
\label{eq:mom}
\end{equation}
\begin{multline}
\left[ h
\left(\rho\frac{u^{2}}{2}+\frac{p}{\gamma-1}\right)
 \right]_t
 +
\left[ h
\left(\rho\frac{u^{2}}{2}+\frac{\gamma p}{\gamma-1} \right) u
\right]_r
\\
  =
   h_{t} \left(\rho\frac{u^{2}}{2} +
   \frac{p}{\gamma-1}\right)
+  \left(  h_{r} -  { n \over r } h  \right)
   \left(\rho\frac{u^{2}}{2}
+\frac{\gamma p}{\gamma-1}\right)   u ,
 \label{eq:energy}
\end{multline}
where $h(t,r)$ and $F(t,r,p\rho^{-\gamma})$ are arbitrary functions.

Equation (\ref{eq:mass}) becomes a homogeneous conservation law if
and only if
\begin{equation*}
F(t,r,z)=r^{n}g(z),
\qquad
z = { p \over \rho^{\gamma} },
\end{equation*}
which for $g \equiv 1$ gives equation (\ref{eq:mass_original}).

Equation
(\ref{eq:mom}) can be a homogeneous conservation law only if $n=0$.
Notice that for $h=r^{n}$ this equation becomes (\ref{eq:mom_original}).
Equation (\ref{eq:energy}) provides a homogeneous conservation law
only for $h=r^{n}$,
it gives equation (\ref{eq:energy_original}).

It should be also noted here that if the overdetermined system defined above is
extended by the condition  $f=0$, then one obtains all possible homogeneous
zero-order conservation laws of the one-dimensional gas dynamics equations.
These conservation laws are discussed in the next section.

%\end{remark}

%\eject

\section{Difference models}

\label{Difference_models}

The first problem in discretization of differential equations is the
choice of a difference mesh. The peculiarity of our approach is that
we add mesh equation(s) into the difference model:
\begin{subequations}
\begin{gather}
F_{i} (z) =0,   \qquad  i = 1,  \ldots , I;
\\
\Omega _j  (z)=0 , \qquad  j = 1,  \ldots , J .
\end{gather}
\end{subequations}
Here the first set of equations  approximates  the underlying differential system
and the second set of equations describes   the difference mesh;
$z$ is a set of difference variables needed for approximation.
As it was shown in \cite{Dorodnitsyn1989, Dorodnitsyn2011}
the invariance of the mesh structure is a necessary condition
for the invariance of the difference model.
The  mesh equations can be presented with the help of difference invariants
or, alternatively,
one can check the invariance of any chosen mesh by means of a certain criterium
(see \cite{Dorodnitsyn1989, Dorodnitsyn2011}).

Symmetries of difference schemes allow one to construct difference
counterparts of the differential conservation laws. The latter provides the
absence of fake sources of energy, impulse,  etc. in difference models
that plays an important role for solutions with big gradients.
%  it time and space.
Moreover, the presence of (local) difference conservation laws
gives a possibility to apply the difference counterpart
of the Gauss-Ostrogradskii theorem~\cite{Samarskii2001}
that leads to global conservation properties of the numerical solutions.

For discretization of the gas dynamics
system~(\ref{Euler_GD_rho}),(\ref{Euler_GD_u}),(\ref{Euler_GD_p}),
which is given in the Eulerian coordinates, the simplest choice
seems to be an orthogonal mesh in $ (t,r) $ plane. As it will be
shown below, this mesh is not invariant with respect to symmetries
which we aim to preserve in the difference models. This
{non}invariance  destroys invariance of difference equations
considered on such a mesh. We will choose another coordinate system
in which one can preserve mesh geometry and, hence, the invariance
of the whole difference model.

% For both differential and difference equations
% symmetries are related to conservation laws.
% preservation of conservation laws
% of the underlying differential equations under discretization % is important
% to reflect the qualitative properties of the solutions
% in numerical simulation.
% For example, it allows to avoid fake sources of
%  energy~\cite{Samarskii2001}.
% In  this section we examine the adiabatic case of
% one-dimensional flows
% for symmetry preservation under discretization.

\subsection{The gas dynamics equations}

In Section~\ref{background}  we considered entropy as one of the dependent
variables.
Since the entropy is conserved along {path}lines only for smooth  solutions
it is appropriate to chose another form of the gas dynamics equations for numerical
modeling.

\subsubsection{Eulerian coordinates }

We start with the equations for the gas dynamics variables $ \rho $, $u$ and $p$:
\begin{subequations}          \label{GD_Eulerian}
\begin{gather}
 \rho   _t   +   u   \rho _r   +  {  \rho   \over r^n  }    ( r^n u )  _r     =   0
 ,
  \\
 u   _t   +   u   u _r   +  {  1 \over  \rho   }    p_r     =   0  ,
  \\
p_t + u p_r + { \gamma p  \over  r^{n}  } (r^{n} u)_{r} = 0    ,
\end{gather}
\end{subequations}
which admits four symmetries for any $n$ and $\gamma$
\begin{multline}        \label{Euler_symmetry}
X_1 =\frac{\partial}{\partial t}   ,
\qquad
X_2 =t \frac{\partial}{\partial t}
 + r \frac{\partial}{\partial r}   ,
\\
X_3 =   2 t \frac{\partial}{\partial t}
  +  r \frac{\partial}{\partial r}
-  u\frac{\partial}{\partial u}
+ 2\rho\frac{\partial}{\partial {\rho}}  ,
\qquad
X_4 =   \rho\frac{\partial}{\partial {\rho}} +  p\frac{\partial}{\partial p}  .
\end{multline}
For  $ n=0$ there are two additional symmetries
\begin{equation}        \label{Euler_symmetries_additional}
X_5 = \frac{\partial}{\partial r}
\qquad
\mbox{and}
\qquad
X_6 = t \frac{\partial}{\partial r}     +  \frac{\partial}{\partial u}   .
\end{equation}
For the special values  $ \gamma _* = { n+3 \over n+1} $
there is one additional symmetry
\begin{equation}        \label{Euler_symmetry_additional}
X_7
= t^2 \frac{\partial}{\partial t}
+ t r  \frac{\partial}{\partial r}
+ (r-tu) \frac{\partial}{\partial u}
-  (n+1)  t \rho \frac{\partial}{\partial {\rho}}
- (n+3)  t p \frac{\partial}{\partial p}    .
\end{equation}

\subsubsection{Conservation laws}

\label{Euler_CL}

System~(\ref{GD_Eulerian}) possesses the following conservation laws.

\begin{enumerate}

\item

General case of $n$ and  $ \gamma $

In the general case we get

\begin{itemize}

\item

Conservation of mass
\begin{equation}      \label{Euler_mass}
\left[ r ^n \rho  \right]   _t   +     \left[ r^n  \rho  u  \right]   _r  = 0
;
\end{equation}

\item

Conservation of energy
\begin{equation}      \label{Euler_energy}
\left[ r  ^n   \left(  \rho   \varepsilon +  { \rho u^2 \over 2 }    \right)
\right]   _t
+   \left[ r^n   \left(  \rho   \varepsilon +  { \rho u^2 \over 2 }    + p
\right)  u  \right]    _r  = 0   ;
\end{equation}

\item

Conservation law
\begin{equation}       \label{Euler_state}
\left[ r ^n \rho   F \left( { p   \over \rho ^{ \gamma } } \right) \right]   _t
+     \left[ r^n  \rho  u    F \left( { p   \over \rho ^{ \gamma } } \right)
\right]   _r  = 0  ,
\end{equation}
where $F$ is a differentiable function.
It holds due to the conservation of mass (\ref{Euler_mass}) and conservation of entropy
along the {path}lines, given by
\begin{equation*}
\left(  { p   \over \rho ^{ \gamma } } \right)   _t
+    u  \left(  { p   \over \rho ^{ \gamma } }  \right)   _r  = 0  .
\end{equation*}

\end{itemize}

\item

Case $ n= 0  $

For the particular case  $ n= 0  $  (plain one-dimensional flows)
we obtain two additional  conservation laws:

\begin{itemize}

\item

Momentum
\begin{equation}        \label{Euler_momentum}
\left[
   \rho    u
\right]   _t
+   \left[
   \rho   u  ^2  +   p
 \right]    _r  = 0    ;
\end{equation}

\item

Motion of the center of mass
\begin{equation}      \label{Euler_center}
\left[
   \rho   ( r - t  u  )
\right]   _t
+   \left[
   \rho   u  ( r  - t u ) - t p
 \right]    _r  = 0     .
\end{equation}

\end{itemize}

\item

Special values of $\gamma _* ={n+3 \over n+1} $

For $\gamma = \gamma_*$ there are two additional  conservation laws
\begin{multline}       \label{Euler_additional_1}
\left[      r^n \left(   2t    \left(  \rho   \varepsilon +  { \rho u^2 \over 2 }
\right)
-   r    \rho u     \right)   \right]   _t
\\
+   \left[    r^n   \left( 2t    \left(  \rho   \varepsilon +  { \rho u^2 \over 2
}    + p   \right)  u
- r     ( \rho u^2 + p   )
 \right) \right]    _r  = 0
\end{multline}
and
\begin{multline}      \label{Euler_additional_2}
\left[   r  ^n      \left(    t^2      \left(  \rho   \varepsilon +  { \rho u^2
\over 2 }    \right)
-  t  r     \rho u
+ { r^2  \over 2 }      \rho
  \right)   \right]   _t
\\
+   \left[ r^n   \left(    t^2     \left(  \rho   \varepsilon +  { \rho u^2 \over
2 }    + p   \right)  u
- t r    ( \rho u^2 + p   )
+ { r^2  \over 2 }     \rho   u
 \right)  \right]    _r  = 0     .
\end{multline}

\end{enumerate}

One can find conservation laws by direct computation
or   by an appropriate reduction of the three-dimensional conservation laws.
Conservation laws of three-dimensional gas dynamics were obtained
in~\cite{Ibragimov2} (see also~\cite{Ibragimov})
with the help of a variational formulation and Noether's theorem
(it requires some assumptions)
and by direct computation in~\cite{Terentev}.
Among the 13  conservation laws of the three-dimensional case
all but one can be integrated over discontinuities~\cite{Terentev}.
The only  conservation law which cannot be integrated over discontinuities gets
reduced to~(\ref{Euler_state})
in the case of one-dimensional flows.
It cannot be integrated over discontinuities because the entropy is  not
conserved for the discontinuous solutions~\cite{Chorin, Landau}.
In~\cite{Ibragimov2, Ibragimov} one can find a symmetry interpretation of the
conservation laws,
i.e.  the correspondence between the conservation laws and Lie point symmetries
of the three-dimensional gas dynamics equations.

% \begin{remark}
% For the gas~(ref)
% there is also  the conservation law
% \begin{equation}       \label{Euler_state}
% \left[ r ^n \rho   F \left( { p   \over \rho ^{ \gamma } } \right) \right]   _t
% +     \left[ r^n  \rho  u    F \left( { p   \over \rho ^{ \gamma } } \right)
%\right]   _r  = 0  ,
% \end{equation}
% where $F$ is a differentiable function.
%  \end{remark}

\subsubsection{Lagrangian  coordinates}

We rewrite the gas dynamics  equations~(\ref{GD_Eulerian})
in the  Lagrangian  coordinates $ (t,s)$ as
\begin{subequations}            \label{gas_dynamics_ for_SP}
\begin{gather}
\left(  { 1 \over \rho }  \right) _t = ( r^n u )   _s     ,
\label{equation_rho}
\\
u _t  +   r^n   p_{s}    = 0    ,
 \label{equation_u}
\\
\varepsilon  _t =   -  p     ( r^n u ) _s    ,
 \label{equation_energy}
\\
r_t =    u    .
 \label{equation_coordinate}
\end{gather}
\end{subequations}
Note that in the Lagrangian coordinates variable $r$ is dependent.
It is  given by  equation~(\ref{equation_coordinate}),
which is included in the system of the gas dynamics  equations,
and the relation
\begin{equation}
r_s ={ 1 \over r^n \rho} .
\end{equation}

From equations~(\ref{gas_dynamics_ for_SP})  it is easy to see that
\begin{equation}     \label{work}
\varepsilon  _t =   -  p   \left(  { 1 \over \rho }  \right) _t    .
\end{equation}
This relation is important for the balance between the specific internal energy
and the specific kinetic  energy.

We rewrite  symmetries~(\ref{Euler_symmetry})
and  additional symmetries~(\ref{Euler_symmetries_additional})
and~(\ref{Euler_symmetry_additional})
in the Lagrangian coordinates.
There are four symmetries in the general case
\begin{multline}        \label{Lagrange_symmetry}
X_1 =\frac{\partial}{\partial t}   ,
\qquad
X_2 =t \frac{\partial}{\partial t}
+ (n+1) s\frac{\partial}{\partial s}
 + r \frac{\partial}{\partial r}   ,
\\
X_3 =   2 t \frac{\partial}{\partial t}
+ (n+3) s\frac{\partial}{\partial s}
-  u\frac{\partial}{\partial u}
+ 2\rho\frac{\partial}{\partial {\rho}}
+  r \frac{\partial}{\partial r} ,
\\
X_4 =  s \frac{\partial}{\partial s}
+ \rho\frac{\partial}{\partial {\rho}} +  p\frac{\partial}{\partial p}  ;
\end{multline}
two additional symmetries for the particular case  $n=0$
\begin{equation}       \label{Lagrange_symmetries_additional}
X_{*,n}  = \frac{\partial}{\partial r}
\qquad
\mbox{and}
\qquad
X_{**,n}  = t \frac{\partial}{\partial r}     +  \frac{\partial}{\partial u}   ;
\end{equation}
and one additional symmetry for the special case $\gamma_*$
\begin{equation}        \label{Lagrange_symmetry_additional}
X_{*, \gamma }
= t^2 \frac{\partial}{\partial t}
+ (r-tu) \frac{\partial}{\partial u}
-  (n+1)  t \rho \frac{\partial}{\partial {\rho}}
- (n+3)  t p \frac{\partial}{\partial p}
+ t r  \frac{\partial}{\partial r}   .
\end{equation}

We also include the translation symmetry for the mass Lagrange coordinate,
which is given by the generator
\begin{equation}      \label{Lagrange_symmetry_s}
X_{0} = \frac{\partial}{\partial s}  .
\end{equation}

\subsubsection{Conservation laws}

\label{Lagrange_CL}

Let us rewrite the conservation laws for the Lagrangian coordinates.
We obtain

\begin{enumerate}

\item

General case of $n$ and $\gamma$

There hold

\begin{itemize}

\item

Conservation of mass
\begin{equation}      \label{Lagrange_mass}
\left[  { 1 \over \rho }  \right]_t
-
 [ r^n  u ]_s
= 0   ;
\end{equation}

\item

Conservation of energy
\begin{equation}      \label{Lagrange_energy}
\left[ \varepsilon   + {1 \over 2} u^2 \right]   _t
+
   [  r^n p   u  ] _s
=   0    ;
\end{equation}

\item

Conservation of entropy along {path}lines
\begin{equation}       \label{Lagrange_state}
\left[ { p   \over \rho ^{ \gamma } } \right]   _t    = 0  .
\end{equation}

\end{itemize}

\item

Case $ n= 0  $

There are additional

\begin{itemize}

\item

Conservation of momentum
\begin{equation}      \label{Lagrange_momentum}
\left[    u \right]    _t
+
  [   p ]  _s
=   0     ;
\end{equation}

\item

Motion of the center of mass
\begin{equation}     \label{Lagrange_center}
\left[ r  -   t  u       \right]    _t
 -    [    t p ]  _s
=   0    .
\end{equation}

\end{itemize}

\item

Special values of $\gamma _* ={ n+3 \over n+1} $

For $ \gamma = \gamma _*  $
there are two additional conservation laws
\begin{equation}       \label{Lagrange_additional_1}
\left[ 2t  \left( \varepsilon   + {1 \over 2} u^2 \right)   - r u \right]    _t
+
  [     r^n    p     ( 2t    u   - r )    ]  _s
=   0
\end{equation}
and
\begin{equation}      \label{Lagrange_additional_2}
\left[ t^2   \left( \varepsilon   + {1 \over 2} u^2 \right)   -   t r u    +  {
r^2 \over 2 }   \right]    _t
+
  [    r^n p (     t^2     u   - t r    ) ]  _s
=   0   .
\end{equation}

\end{enumerate}

% \begin{remark}
% For the gas~(ref) we also have  the conservation law
% \begin{equation}       \label{Lagrange_state}
% \left[ { p   \over \rho ^{ \gamma } } \right]   _t    = 0  ,
% \end{equation}
% which is the analog of~(\ref{Euler_state}).
% \end{remark}

%\eject

\subsection{The numerical schemes}

In this section we consider  numerical schemes and their symmetries.
Besides, our goal is  to construct schemes which have difference conservation laws
analogous to the conservation laws of the underlying differential system.
We restrict ourselves by the homogenous conservation laws.
%  for $n=1,2$.

\subsubsection{Invariance and Eulerian coordinates for $ n=0$ }

 For discretization of the gas dynamics
system~(\ref{GD_Eulerian}),
which is given in Eulerian coordinates, the simplest choice seems to
be an orthogonal mesh in $ (t,r) $ plane. However, this mesh is not
invariant that destroys invariance of difference equations
considered on such mesh. Indeed, as it was shown
in~\cite{Dorodnitsyn1989,  Dorodnitsyn2011} the necessary condition
for a mesh to preserve its orthogonality under a group
transformation generated by the operator
\begin{equation}
X=
\xi ^t  {\partial \over \partial t }
+  \xi ^r  {\partial \over \partial r }
+ \ldots
\end{equation}
is the following:
\begin{equation}  \label{orthogonality}
D_{+h} (\xi^t) = -D_{+\tau} (\xi^x) ,
\end{equation}
where $D_{+h}$ and $D_{+\tau}$ are the operators of difference
differentiation in $r$ and $t$ directions respectively.

System~(\ref{GD_Eulerian})    admits
the $6$-parameter Lie symmetry group of point
transformations  that corresponds to the Lie algebra of
infinitesimal operators  (\ref{Euler_symmetry})
and      (\ref{Euler_symmetries_additional}).
In the special case $\gamma = 3$
there  is one more symmetry   (\ref{Euler_symmetry_additional}).

It is easy to see that the  Galilean transformation  given by the
operator $X_{**,n}$ does not satisfy the
criterion~(\ref{orthogonality}). The same is true for $X_{*, \gamma}
$. It means that one  should look for an invariant moving mesh in
the Eulerian   coordinates.

To obtain an invariant moving mesh we chose   the following  difference stencil with
two time layers:

\begin{itemize}

\item

independent variables:
$$
 t = t_j  ,
\quad
 \hat{t}=   t _{j+1} ;
\quad
 r = r_{i} ^{j}  ,
\quad
r_+   = r_{i+1} ^{j} ,
\quad
 \hat{r} = r_{i} ^{j+1},
\quad
 \hat{r}_+ = r_{i+1} ^{j+1};
$$

\item

dependent variables in the nodes of the mesh (the same notation as for $r$):
% $  u   $,   $ u_+$,  $ \hat{u}$,   $\hat{u}_+$,
$$
  u   ,
\quad
u_+    ,
\quad
 \hat{u} ,
\quad
 \hat{u}_+ ;
\quad
  \rho      ,
\quad
 \rho   _-    ,
\quad
 \hat{\rho} ,
\quad
 \hat{\rho}_-   ;
% $  \rho  = \rho  _{i} ^{j}   $,   $  \rho _-   =  \rho  _{i-1} ^{j}   $,
% $ \hat{\rho}  = \rho  _{i} ^{j+1}   $,   $\hat{\rho}_-   =  \rho  _{i-1|} ^{j+1}$,
 \quad
p    ,
\quad
p _-  ,
\quad
\hat{p} ,
\quad
 \hat{p}_-   .
$$
\end{itemize}
Then, we find the finite-difference invariants for symmetries~(\ref{Euler_symmetry})
and      (\ref{Euler_symmetries_additional})
as solutions of the system of linear equations
\begin{equation}
X _i   I  (   t   ,    \hat{t}  ,
 r  ,   r_+    ,   \hat{r}  , \hat{r}_+ ,
 \ldots  ,
 p      ,  p  _-    , \hat{p} ,  \hat{p}_-    )  = 0
\end{equation}
for the considered symmetries.
Here we assume that the operator is prolonged for all variables of the
stencil~\cite{Dorodnitsyn2011}.
There are 12 functionally independent invariants
\begin{equation*}
\frac{\hat{h}_+}{h_+}  ,
\qquad
\frac{\tau}{h_+}   \sqrt{\frac{p}{\rho}}  ,
\qquad
\sqrt{\frac{\rho}{p}}  \left( \frac{ \hat{r} - r } { \tau } - u \right)  ,
\end{equation*}
\begin{equation*}
\sqrt{\frac{\rho}{p}}(u_+ -u) ,
\qquad
\sqrt{\frac{\rho}{p}}(\hat{u}-u) ,
\qquad
\sqrt{\frac{\rho}{p}}(\hat{u}_+ - \hat{u}) ,
\end{equation*}
\begin{equation*}
\frac{p_+}{p}  ,
\qquad
\frac{\hat{p}}{p}  ,
\qquad
\frac{\hat{p}_+}{\hat{p}}  ,
\qquad
\frac{\hat{\rho}}{\rho}  ,
\qquad
\frac{\hat{\rho}_+}{\hat{\rho}}  ,
\quad
\frac{{\rho_+}}{\hat{\rho}}  ,
\end{equation*}
where  $ \tau =  \hat{t}   -   t   $,
$ h_+   =  r _ +    -  r   $
and
$ \hat{h} _+   =   \hat{r}  _ +    -  \hat{r}   $.

Notice, that the only one difference invariant contains the value
$\hat{r}$. This invariant suggests, for example,
 an  invariant moving mesh given by
\begin{equation*}
\sqrt{\frac{\rho}{p}}  \left( \frac{ \hat{r} - r } { \tau } - u  \right) = 0
\end{equation*}
or,  equivalently,
\begin{equation}
\frac{ \hat{r} - r } { \tau } = u .
\end{equation}
In the continuous limit it corresponds to the evolution of the spacial  variable $r$
given as
\begin{equation}
\frac{dr}{dt} =u  .
\end{equation}
Thus, we arrive at choosing the mass Lagrangian coordinates with the
operator of differentiation with respect to $t$
\begin{equation*}
D_t ^L = D_t ^E + u D_r .
\end{equation*}

% Below we introduce a  scheme, which is invariant with respect to
% symmetries of 1+1 D gas dynamics in the case  $ \gamma  \neq  3$.

\subsubsection{Notations}

We introduce the mesh for the mass Lagrangian coordinate $s$:
\begin{equation}
h^s = s_{i+1} - s _i
\qquad
\mbox{and}
\qquad
h^s _- = s_{i} - s _{i-1}   .
\end{equation}
Generally, the spacing can be nonuniform.
For simplicity we use a uniform mesh  $ h^s = h^s _-  $.

For time we consider the mesh with points $t_j$.
Since we consider the schemes with two time  layers we denote the time step as
$\tau$.
Of course, we can consider nonuniform time meshes  with {step}lengths $ \tau _j =
t_{j+1} - t_j $.

Now the operators have the form
\begin{equation}
X=
\xi ^t  {\partial \over \partial t }
+  \xi ^s  {\partial \over \partial s }
+ \ldots
\end{equation}
and the criterium of invariant orthogonality
\begin{equation}  \label{orthogonality_new}
D_{+h_s} (\xi^t) = -D_{+\tau} (\xi^s)
\end{equation}
holds for all considered
symmetries~(\ref{Lagrange_symmetry}),
(\ref{Lagrange_symmetries_additional}),
   (\ref{Lagrange_symmetry_additional})
   and~(\ref{Lagrange_symmetry_s}).
Here $D_{+h_s}$ and $D_{+\tau}$ are the operators of difference
differentiation in $s$ and $t$ directions,  respectively.

We split the dependent variables into kinematic and thermodynamic.
The kinematic variables $u$  and $r$ are prescribed to the nodes.
For example, for $u$ we have
\begin{equation*}
u =  u_ i  ^j ,
\qquad
u_+ =  u_{i+1} ^j  ,
\qquad
\hat{u}  =  u_ i  ^{j+1}  ,
\qquad
\hat{u} _+ =  u_{i+1} ^{j+1}    .
\end{equation*}
The thermodynamic variables    $\rho$ and $p$ are taken in the midpoints as
\begin{equation*}
\rho_-   =  \rho  _ {i-1/2}  ^j  ,
\qquad
\rho   =  \rho  _ {i+1/2}  ^j  ,
\qquad
\rho  _+ =  \rho  _ {i+3/2} ^j  .
\end{equation*}

To describe the scheme we need the time and spatial derivatives
\begin{equation*}
u_t = { \hat{u}   -  u   \over  \tau  } ,
\qquad
u_s = { u_{i+1}  ^j   - u _i ^j    \over  s_{i+1} - s_i   }
=  { u_{+}   - u   \over   h ^s  }  .
% \qquad
% u_{\bar{s}} = { u _i ^j    - u_{i-1}  ^j  \over  s_i   - s _{i-1} }
% =  { u   - u_-   \over   h ^s    }   ,
\qquad
 p_{\bar{s}} = { p _{ i+1/2 }  ^j -   p _{ i-1/2 }  ^j \over  {1\over 2} ( s_{i+1}
 - s _{i-1}  )      }
= { p  -   p _- \over  h_s       }
% \qquad
% p_{s} = { p _{ i+3/2 } ^j  -   p _{ i+1/2 }  ^j \over   h^s       }  .
\end{equation*}
and weighted values defined  as
% \begin{equation}
%   p_{*}    = (  p_{*} ) _ i   ^j   = {   p _{ i-1/2 }  ^j   +    p _{ i+1/2 }  ^j
%\over   2     }  ,
% \end{equation}
\begin{equation*}
 y    ^{(\alpha)}    =  \alpha   \hat{y} + ( 1 - \alpha ) y   ,
\qquad
0 \leq \alpha \leq 1   .
\end{equation*}

%\eject

\subsubsection{The Samarskii--Popov  scheme}

In~\cite{Popov} (see also~\cite{Samarskii}) the authors introduced a conservative
scheme
for plain one-dimensional flows ($n=0$).
It was generalized to the other one-dimensional flows  ($n=1, 2$)
 in~\cite{Samarskii}.
This scheme is  a discretization of the equations~(\ref{gas_dynamics_ for_SP})
\begin{subequations}      \label{Samarskii_Popov_scheme}
\begin{gather}
\left(  { 1 \over \rho }  \right) _t = ( R u  ^{(0.5)}   )_s    ,
   \label{equation_discrete_mass}
\\
u _t =   -   R   p_{\bar{s}}    ^{(\alpha)}   ,
\label{equation_discrete_velocity}
\\
\varepsilon  _t =   -  p ^{(\alpha)} (   R  u  ^{(0.5)} ) _s    ,
  \label{equation_discrete_energy}
\\
r_t =    u  ^{(0.5)}       ,
 \label{equation_discrete_coordinate}
\end{gather}
\end{subequations}
where $R$  is a discretization of $r^n$ chosen as

\begin{equation*}
R = { \hat{r} ^{n+1} -    {r} ^{n+1}  \over ( n+1 ) ( \hat{r}  - r )  }
=
\left\{
\begin{array}{ll}
{\displaystyle   1  }  , &   n  =  0 ; \\
 &  \\
{\displaystyle   { \hat{r} +  r  \over 2 }   }  , &   n  =  1 ; \\
 &  \\
 {\displaystyle   { \hat{r}^2  +   \hat{r} r  +  r^2  \over 3 }   }      , &   n  =
 2 . \\
\end{array}
\right.
\end{equation*}

% This schemes stands as a generalization of a conservative scheme
%  for plane one-dimensional flows
%  suggested by the same authors in~\cite{Popov, Samarskii}.

Scheme  (\ref{Samarskii_Popov_scheme})
has  four equations for five variables
$\rho$, $u$,  $\varepsilon$, $r$ and $p$.
It should be supplemented by a discrete equation of state,
a discrete analog of  (\ref{general_equation_of_state}).
For example, it can be taken in the same form that means
\begin{equation}    \label{discrete_state}
\varepsilon   _{i+1/2} ^j     =   \varepsilon   ( \rho     _{i+1/2} ^j , p    _{i+1/2} ^j ).
\end{equation}

\subsubsection{Properties of the Samarskii--Popov scheme}

For a polytropic gas
scheme~(\ref{Samarskii_Popov_scheme}),(\ref{discrete_state}) is invariant
with respect to the symmetries~(\ref{Lagrange_symmetry})
and~(\ref{Lagrange_symmetry_s})
corresponding to the general case.
For  $ n=0$ it is also invariant to
symmetries~(\ref{Lagrange_symmetries_additional}).
The scheme  is not invariant for the additional
symmetry~(\ref{Lagrange_symmetry_additional}),
which exists for the special values  $\gamma_*$.

Let us review important properties of the scheme.
It possesses many qualitative properties of the underlying differential equations.
For any equation of state   $\varepsilon =  \varepsilon  ( \rho , p)$,
i.e.  not only for the polytropic  gas~(\ref{equation_of_state}),
this scheme has the following   conservation laws:

\begin{itemize}

\item

Conservation of mass

\begin{equation}         \label{scheme_mass}
\left(  { 1 \over \rho }  \right) _t = ( R u  ^{(0.5)}   )_s   ;
\end{equation}

\item

Conservation of energy
\begin{equation}         \label{scheme_energy}
\left[ \varepsilon   + {  u^2 + u_+ ^2    \over 4}       \right]   _t
+
   [ R   p _{*} ^{(\alpha)}    u ^{(0.5)}  ] _s
=   0  ,
\end{equation}
where
\begin{equation*}
  p_{*}  ^{(\alpha)}  = (  p_{*} ) _ i   ^{(\alpha)}
= {   p _{ i-1/2 }  ^{(\alpha)}   +    p _{ i+1/2 }  ^{(\alpha)}    \over   2
}     .
\end{equation*}

\end{itemize}
For $n=0$ there are two additional conservation laws:

\begin{itemize}

\item
Conservation of momentum
\begin{equation}
 \left[ u       \right]   _t
 +
    [    p  ^{(\alpha)} ] _s
 =   0   ;
 \end{equation}

\item

Motion of the  center of mass
\begin{equation}
 \left[ r - t u       \right]   _t
 -
    [   t  ^{(0.5)}   p  ^{(\alpha)}  ] _s
 =   0   .
 \end{equation}

\end{itemize}
These conservation laws correspond to~(\ref{Lagrange_mass}),
(\ref{Lagrange_energy}),  (\ref{Lagrange_momentum})  and~(\ref{Lagrange_center}).
There are no discrete conservation laws
corresponding to~(\ref{Lagrange_additional_1}) and~(\ref{Lagrange_additional_2}),
which hold for the special values of $\gamma_*$.

%\eject

\begin{remark}   \label{more_conservation}
Modifying the equation of state~(\ref{equation_of_state}),
it is possible to achieve
conservation of the conservation laws~(\ref{Lagrange_additional_1})
and~(\ref{Lagrange_additional_2}),
which hold for $ \gamma _* =  { n+3 \over n+1   } $,
 under discretization.
We refer to~\cite{Korobitsyn} for the case $n =0 $
and to~\cite{Kozlov} for the generalization to  $n =1, 2$.

% For $ \gamma _* =  { n+3 \over n+1   } $
% we are interested to preserve the additional conservation
% laws~(\ref{Lagrange_additional_1})
% and~(\ref{Lagrange_additional_2}).
Scheme (\ref{Samarskii_Popov_scheme})
consists of four equations for five variables $ \rho$, $u$, $p$, $\varepsilon$ and $r$.
 We will not impose the discrete equation of state   (\ref{discrete_state}).
The freedom to choose a discretization of the equation of state will be used
to impose open additional conservation law.
% We can use the remaining freedom to chose the discrete equation of state.
Let us look for an equation of state which gives us the following difference analog
of the additional conservation law~(\ref{Lagrange_additional_1}):
\begin{equation}          \label{scheme_additional_1}
\left[
2t  \left( \varepsilon   + { < u^2 >     \over 2}       \right)
-  < r u >
\right]    _t
+
  [   R   p  _{*} ^{(\alpha)}     ( 2t ^{(0.5)}    u  ^{(0.5)}     -    r ^{(0.5)}
  )     ]  _s
=   0    ,
\end{equation}
where we use  a special notation for the average value of two function values
taken in  the neighbouring nodes of the same time layer
\begin{equation*}
< f (u,r) > = { f (u,r)  + f (u_+,r_+)  \over 2}   .
\end{equation*}

It leads to  the following specific  internal energy equation
\begin{equation}      \label{energy_definition}
\varepsilon   ^{(0.5)}  =
{  p ^{(\alpha)}   \over   \gamma -1  }   \left(   { 1\over \rho}   \right)
^{(0.5)}
- {  \tau ^2  \over 8}       <  (  u_t  ) ^2 >
+ { 1 \over 2 } p ^{(\alpha)}
\left[   r ^{(0.5)}   R   -     (   r  ^{n+1}  ) ^{(0.5)}   \right]  _s    .
\end{equation}
We will take it as the {\it discrete equation of state}, which
approximates~(\ref{equation_of_state}).
% with the same order as for other equations.

In this case we also get a difference analog
of the second additional conservation law~(\ref{Lagrange_additional_2}) as
\begin{multline}        \label{scheme_additional_2}
\left[
t^2   \left( \varepsilon   + {  < u^2 >     \over 2}       \right)
-   t  < r u >
+  {  < r^2 >  \over 2 }
+  { \tau ^2 \over 8}      { < u^2 > }
\right]    _t
\\
\\
+
  [     R      p  _{*} ^{(\alpha)}
(  (  t^2 )  ^{(0.5)}       u  ^{(0.5)}     -  t   ^{(0.5)}    r    ^{(0.5)}    )
]  _s
=   0   .
\end{multline}
Note that it has a correcting term  $ { \tau ^2  \over 8}     { < u^2 >  }   $,
which disappears in the continuous limit.

Thus, we obtained difference scheme~(\ref{Samarskii_Popov_scheme})
supplemented by discrete state equation~(\ref{energy_definition}).
In this scheme the pressure values  $p$ and $ \hat{p}$
appear only as a weighted value $  p^{(\alpha)} $,
i.e. $ \alpha$  has no longer meaning of a parameter.
We can consider this value
as the  pressure in the midpoint of the cell $ (  t  _{j+1/2}  ,   s  _{i+1/2} )  $,
i.e.  for  $\alpha = 0.5$.

\end{remark}

%\eject

% Balance  of specific energies

The scheme holds  a discrete counterpart of the relation~(\ref{work}), namely
\begin{equation}     \label{discrete_work}
\varepsilon  _t =   -  p ^{(\alpha)}   \left(  { 1 \over \rho }  \right) _t  .
\end{equation}
This is an important supplement to the conservation of total
energy~(\ref{scheme_energy}),
which provides the balance of the specific internal energy and the specific
kinematic energy.

% {Non}conservation  of  entropy

In case of a  polytropic gas the equations of gas dynamics
hold the conservation of entropy~(\ref{Lagrange_state})
 along {path}lines (for smooth solutions).
There is no such property for the scheme~(\ref{Samarskii_Popov_scheme}).
However, the scheme holds the relation
\begin{equation}
{ \Delta   p  \over   p ^{(\alpha)}     }
= \gamma
{ \Delta   {\rho}    \over   \rho   ^{(\alpha)}      }    ,
\qquad
 \Delta   p  = \hat{p} - p ,
\qquad
 \Delta   \rho   = \hat{\rho} - \rho
\end{equation}
that approximates~(\ref{Lagrange_state})   presented with the help of differentials
\begin{equation}
{ d p     \over   p    }
= \gamma
{   d {\rho}   \over   \rho   }     .
\end{equation}

%\eject

\subsection{Invariance of difference schemes}

In this point
we show how to construct invariant schemes with the help of finite-difference
invariants.
Scheme~(\ref{Samarskii_Popov_scheme})
can be expressed in terms of invariants for general case of $\gamma$.
Its modification described in Remark~\ref{more_conservation}
possesses the additional conservation laws which hold for the  special values $\gamma_*$.
However, it is not invariant with respect to the additional symmetry  $ X_{*, \gamma}$
which exists for  these special values.
For the special values $\gamma_*$  invariant schemes are constructed.
% Here we assume $n\neq 0 $.
The case   $n =  0 $ was reported in~\cite{DKM1}.

\subsubsection{General  case  $n\neq 0$, $\gamma \neq { n +3 \over n+1} $}

We chose an orthogonal mesh in the  Lagrangian coordinates
and a stencil with the following variables

\begin{itemize}

\item

independent variables:
\begin{equation*}
t = t_j  ,
 \quad
\hat{t} = t _{j+1} ,
 \quad
s = s_i  ,
\quad
s_+  =   s_{i+1}  ,
\quad
 s_-  =    s_{i-1};
\end{equation*}

\item

kinematic variables in the nodes:
\begin{equation*}
  u = u_{i} ^{j}  ,
\quad
u_+   = u_{i+1} ^{j} ,
\quad
 \hat{u} = u_{i} ^{j+1},
\quad
 \hat{u}_+ = u_{i+1} ^{j+1} ,
 \quad
 r   ,
\quad
r _+  ,
\quad
 \hat{r} ,
\quad
 \hat{r} _+ ;
 \end{equation*}

\item

thermodynamic variables in the midpoints:
\begin{equation*}
  \rho    =  \rho   _{i} ^{j}  ,
\quad
 \rho   _-   =  \rho   _{i-1} ^{j} ,
\quad
 \hat{\rho} =  \rho   _{i} ^{j+1},
\quad
 \hat{\rho}_- =  \rho   _{i-1} ^{j+1} ,
% $  \rho  = \rho  _{i} ^{j}   $,   $  \rho _-   =  \rho  _{i-1} ^{j}   $,
% $ \hat{\rho}  = \rho  _{i} ^{j+1}   $,   $\hat{\rho}_-   =  \rho  _{i-1|} ^{j+1}$,
 \quad
p    ,
\quad
p _-  ,
\quad
\hat{p} ,
\quad
 \hat{p}_-    .
\end{equation*}
\end{itemize}

\bigskip
\bigskip

For these 21 stencil variables we find   16  = 21 - 5  invariants
of the symmetries~(\ref{Lagrange_symmetry}) and~(\ref{Lagrange_symmetry_s}):
\begin{equation*}
I_{1} = { h^s _-   \over h^s }    ,
 \quad
I_{2} =    { \rho   r^{n+1}   \over h^s }   ,
 \quad
I_{3} =    {  \tau   \over   h^s }  r^n   \sqrt{ \rho p  }   ,
\quad
I_{4} = \frac{\tau u }{r}  ,
\end{equation*}
\begin{equation*}
I_{5} = \frac{u _+}{u}  ,
\quad
I_{6} = \frac{\hat{u}}{u}  ,
\quad
I_{7} = \frac{\hat{u}_+}{\hat{u}}  ,
\quad
I_{8} = \frac{r_+}{r}  ,
\quad
I_{9} = \frac{\hat{r}}{r}  ,
\quad
I_{10} = \frac{\hat{r}_+}{\hat{r}}    ,
\end{equation*}
\begin{equation*}
I_{11} = \frac{\rho_-}{\rho}  ,
\quad
I_{12} = \frac{\hat{\rho}}{\rho}  ,
\quad
I_{13} = \frac{\hat{\rho}_-}{\hat{\rho}}  ,
\quad
I_{14} = \frac{p_-}{p}  ,
\quad
I_{15} = \frac{\hat{p}}{p}  ,
\quad
I_{16} = \frac{\hat{p}_-}{\hat{p}}    .
\end{equation*}

\bigskip
\bigskip

The scheme~(\ref {Samarskii_Popov_scheme})  is invariant with respect to the
considered symmetries
and can be expressed in terms of the invariants as
\begin{subequations}
\begin{gather}
 I_{12} - 1
=
  I_{2}  I_{4}
\left(
{     (  I_{9}  I_{10}  ) ^{n+1} -1    \over  (n+1)   (  I_{9}   I_{10}  - 1  )  }
{    I_{6}  I_{7}   + I_{5}   \over 2 }
-
{   I_{9}  ^{n+1}  - 1     \over  (n+1)     ( I_{9}   - 1 )   }
{    I_{6}   + 1   \over 2 }
\right)   ,
\\
 I_{6}  - 1
=
 -   {  I_{3}  ^2    \over   I_{2}   I_{4}   }
   {   I_{9}  ^{n+1}  - 1     \over  (n+1)     ( I_{9}   - 1 )   }
\left(
\alpha   I_{15}  ( 1 -  I_{16}  )  + ( 1 - \alpha  )  ( 1 -  I_{14}  )
\right)    ,
\\
\begin{array}{c}
\displaystyle
{ 1 \over \gamma  -1 }
\left(
{     I_{15}  \over   I_{12}  }  - 1
\right)
=
-  I_{2}   I_{4}
 \left(
\alpha   I_{15}   + ( 1 - \alpha  )
 \right)
\\
\displaystyle
\times
\left(
{     (  I_{9}  I_{10}  ) ^{n+1} -1    \over  (n+1)   (  I_{9}   I_{10}  - 1  )  }
{    I_{6}  I_{7}   + I_{5}   \over 2 }
-
{   I_{9}  ^{n+1}  - 1     \over  (n+1)     ( I_{9}   - 1 )   }
{    I_{6}   + 1   \over 2 }
\right)     ,
\end{array}
\\
I_{9} - 1
=
 { 1 \over 2 }
I_{4}    ( 1 + I_{7}  )   .
\end{gather}
\end{subequations}

\subsubsection{Special  case $n =0$,
   $\gamma \neq {n+3 \over n+1} $}

In the space of 21 stencil variables there are 14 invariants for
7 symmetries~(\ref{Lagrange_symmetry}),
(\ref{Euler_symmetries_additional}),  (\ref{Lagrange_symmetry_s}):
\begin{equation*}
I_{1} = { h^s _-   \over h^s }    ,
 \quad
I_{2} = \frac{\tau}{h^s}  \sqrt{ {\rho}{p} } ,
 \quad
I_{3} = \sqrt{ \frac{\rho}{p} } \left( \frac{ \hat{r} - r } { \tau } - u  \right),
\quad
I_{4} = \sqrt{ \frac{{\rho}}{{p}} }   \left( \frac{ \hat{r} - r } { \tau } - \hat{u}
\right),
\end{equation*}
\begin{equation*}
I_{5} = \sqrt{ \frac{\rho}{p} }(u_+ -u),
\quad
I_{6} = \sqrt{\frac{\rho}{p}}(\hat{u}_+ - \hat{u}),
\quad
I_{7} = {  \rho ( r_+ - r)   \over h^s }   ,
\quad
I_{8} = {  \hat{\rho} ( \hat{r}_+ - \hat{r} )   \over h^s }   ,
\end{equation*}
\begin{equation*}
I_{9} = \frac{\rho_-}{\rho}  ,
\quad
I_{10} = \frac{\hat{\rho}}{\rho}  ,
\quad
I_{11} = \frac{\hat{\rho}_-}{\hat{\rho}}  ,
\quad
I_{12} = \frac{p_-}{p}  ,
\quad
I_{13} = \frac{\hat{p}}{p}  ,
\quad
I_{14} = \frac{\hat{p}_-}{\hat{p}}    .
\end{equation*}

One can find the scheme
(\ref{Samarskii_Popov_scheme}) for $ n=0 $ approximating the gas dynamics system
(\ref{gas_dynamics_ for_SP}) with the help of these invariants as
\begin{subequations}
\begin{gather}
{ 1 \over I_{10} }  - 1 =   I_{2}    {   I_{5} +  I_{6}   \over 2 }   ,
\\
I_{3}  -  I_{4}  =  - I_{2}   \left(
\alpha \left(   I_{13}  -    { I_{13}  I_{14} }      \right)
+ ( 1 - \alpha ) ( 1  - I_{12} )
\right)   ,
\\
{ 1 \over  \gamma -1 } \left( {  I_{13}  \over  I_{10}  }  - 1 \right)
= -   I_{2}    (     \alpha   I_{13}    +   ( 1 - \alpha ))     {   I_{5} +  I_{6}
\over 2 }    ,
\\
 I_{3} +  I_{4}   = 0 .
\end{gather}
\end{subequations}

% Reference on our first paper~\cite{DKM1}.

%\eject

\subsubsection{Special  case $ n\neq 0$, $\gamma _* = {n+3 \over n+1} $}

We use the same mesh and stencil as for the general case of $\gamma$.
Due to the additional symmetry~(\ref{Lagrange_symmetry_additional})
we get one invariant less.
We obtain the following finite-difference invariants:
\begin{equation*}
J_{1} = { h^s _-   \over h^s }    ,
\quad
J_{2} =  { \rho r^{n+1} \over h_s   }  ,
\quad
J_{3} = {  \hat{\rho}    \hat{r}   ^{n+1}    \over h^s }   ,
\quad
J_{4} = { \tau   r ^n  \over h_s }
\rho  ^{  {1\over 2 } - { 1\over  n+1 }  }
 \hat{\rho} ^{ 1 \over n + 1}
p    ^{  1\over 2 }  ,
\quad
J_{5} = \frac{\hat{p}}{p}   \left(  \frac{\rho}{\hat{\rho}} \right)^{n+3 \over n+1}
,
\end{equation*}
\begin{equation*}
J_{6} = {  r + \tau u \over \hat{r} }
\quad
J_{7} = {  r_+  + \tau u_+  \over \hat{r}_+  } ,
\quad
J_{8} = {  \hat{r}  - \tau \hat{u}  \over r  }     ,
\quad
J_{9} =  {  \hat{r} _+  - \tau \hat{u} _+   \over r _+  }    ,
\end{equation*}
\begin{equation*}
J_{10} = { r_+ \over r }
\quad
J_{11} = { \hat{r}_+ \over \hat{r} } ,
\quad
J_{12} = \frac{\rho_-}{\rho}  ,
\quad
J_{13} = \frac{\hat{\rho}_-}{\hat{\rho}}  ,
\quad
J_{14} = \frac{p_-}{p}  ,
\quad
J_{15} = \frac{\hat{p}_-}{\hat{p}}    .
\end{equation*}

Using these invariants,  we suggest an   invariant scheme
\begin{subequations}   \label{explicit_scheme}
\begin{gather}
   \hat{\rho}  (  \hat{r} _+ ^{n+1}    - \hat{r}  ^{n+1} )  =  {\rho}   (   {r} _+
   ^{n+1}    - r  ^{n+1} )     ,
 \label{explicit_scheme_rho}
\\
{ \frac{\hat{u} - u}{\tau}  } =
-
\left(
{ \frac{\hat{\rho}}{\rho}}
 \right)  ^{ 2  \over  n+1 }
 r^n  \    { \frac{p   - p_- }{h ^s} }     ,
\label{explicit_scheme_u}
\\
 \frac{\hat{p}}{\hat{\rho} ^{n+3 \over n+1} } =  \frac{p}{{\rho}  ^{n+3 \over n+1} }
\label{explicit_scheme_p}  ,
\\
 \frac{ \hat{r} - r }{ \tau } =  u
\label{explicit_scheme_x}  ,
\end{gather}
\end{subequations}
which allows explicit computations.
It is expressed in terms of the invariants as
\begin{subequations}
\begin{gather}
J_3 ( J_{11}^ {n+1} - 1 )  = J_2 ( J_{10}^ {n+1} - 1 )   ,
\\
J_8 - 1
=
{  J_4  ^2 \over  J_2   }
(   1 - J _{14} )  ,
\\
J_5 =  1   ,
\\
J_6 = 0    .
\end{gather}
\end{subequations}

In addition to the invariance the scheme~(\ref{explicit_scheme})
possesses conservation of mass, given by equation~(\ref{explicit_scheme_rho}),
and conservation of the entropy along {path}lines, given
by~(\ref{explicit_scheme_p}).

We remark that the conservation of mass property can be rewritten as
\begin{equation}
  {1\over \tau  } \left( {  1 \over \hat{\rho}  }   -     {  1 \over {\rho}  }
  \right)
 =  { R_+ u _+ - R u   \over h^s }
\qquad
\mbox{or}
\qquad
 \left(  {  1 \over {\rho}  }  \right) _ t = ( R u ) _s
\end{equation}
with
\begin{equation*}
h_s
=     \hat{\rho}  { \hat{r} _+ ^{n+1}    - \hat{r}  ^{n+1}  \over n+1 }
=  {\rho}   {    {r} _+ ^{n+1}    - r  ^{n+1} \over n+1}    .
\end{equation*}

 \subsubsection{Special  case  $n =0$,  $ \gamma  = 3 $}

% For the case ${\gamma = 3 }$ we can suggest invariant schemes,
% but there are difficulties in  finding invariant schemes  with conservation laws.

In comparison to the case $ n= 0 $, ${\gamma \neq  3 }$
we have one more symmetry, namely~(\ref{Lagrange_symmetry_additional}).
Therefore, we get one invariant less. There are  13 invariants:
\begin{equation*}
J_{1} = { h^s _-   \over h^s }    ,
 \quad
J_{2} = \frac{\tau}{h^s}   ( {\rho}{p} \hat{\rho} \hat{p}  )   ^{1 \over 4}  ,
 \quad
J_{3} = \sqrt{ \frac{\rho}{p} } \left( \frac{ \hat{r} - r } { \tau } - u  \right),
\quad
J_{4} = \sqrt{ \frac{\hat{\rho}}{\hat{p}} }   \left( \frac{ \hat{r} - r } { \tau } -
\hat{u} \right),
\end{equation*}
\begin{equation*}
J_{5} = \sqrt{ \frac{\rho}{p} }
\left( { \frac{h_+}{\tau} } +  u_+ -u \right),
\quad
J_{6} = \sqrt{ \frac{\hat{\rho}}{\hat{p}} }
\left( - { \frac{\hat{h}_+}{\tau} } +  \hat{u}_+ -\hat{u} \right),
\quad
J_{7} = {  \rho ( r_+ - r)   \over h^s }   ,
\end{equation*}
\begin{equation*}
J_{8} = {  \hat{\rho} ( \hat{r}_+ - \hat{r} )   \over h^s }   ,
\quad
J_{9} = \frac{\hat{p}}{p}   \left(  \frac{\rho}{\hat{\rho}} \right)^3    ,
\quad
J_{10} = \frac{\rho_-}{\rho}  ,
\quad
J_{11} = \frac{\hat{\rho}_-}{\hat{\rho}}  ,
\quad
J_{12} = \frac{p_-}{p}  ,
\quad
J_{13} = \frac{\hat{p}_-}{\hat{p}}    .
\end{equation*}

There are many possibilities to approximate
the gas dynamics system~(\ref{gas_dynamics_ for_SP}),(\ref{Lagrange_state})
 with the help of these invariants.
We propose the following explicit invariant scheme:
\begin{subequations}   \label{explicit_scheme_n}
\begin{gather}
   \hat{\rho}  (  \hat{r} _+   - \hat{r} )  =  {\rho}   (   {r} _+   - r )     ,
 \label{explicit_scheme_rho_n}
\\
{ \frac{\hat{u} - u}{\tau}  } =
-
\left(
{ \frac{\hat{\rho}}{\rho}}
 \right)  ^2
{ \frac{p   - p_- }{h ^s} }     ,
\label{explicit_scheme_u_n}
\\
 \frac{\hat{p}}{\hat{\rho} ^3} =  \frac{p}{{\rho} ^3}
\label{explicit_scheme_p_n}  ,
\\
 \frac{ \hat{r} - r }{ \tau } =  u
\label{explicit_scheme_x_n}  .
\end{gather}
\end{subequations}
In term of the invariants this scheme is written as
\begin{subequations}
\begin{gather}
J_{7} =    J_{8}      ,
\\
 J_{4}  =     J_{2}    J_{9} ^{-3/4}      (    1 - J_{12}  )   ,
\\
J_{9} = 1   ,
\\
J_{3} = 0     .
\end{gather}
\end{subequations}
The scheme conserves the entropy, or  $S$,  along the  streamlines
and possesses conservation of mass
(\ref{explicit_scheme_rho}).
Note that the first equation can be rewritten as
\begin{equation}
  {1\over \tau  } \left( {  1 \over \hat{\rho}  }   -     {  1 \over {\rho}  }
  \right)
 =  { u _+ - u   \over h^s }     .
\end{equation}
We remark that implicit invariant schemes are  also possible.

% Reference on our first paper~\cite{DKM1}.

\section{Conclusion}

\label{Concluding}

In the paper we examined one-dimensional flows of a polytropic gas
and their Lie point symmetry properties. By the one-dimensional
flows we mean plain one-dimensional flows, the gas dynamics flows
with radial symmetry and the gas dynamics flows  with spherical
symmetry. There was performed Lie group classification of the gas
dynamics equations reduced to a single second-order PDE in the
Lagrangian coordinates. The entropy function was a parameter of the
classification. Four cases were identified. In the general case
there are conservation laws of mass and energy. For the special
cases there were found additional conservation laws. The
conservation laws obtained for the second-order PDE were later
rewritten for the gas dynamics variables. They were also transformed
from the Lagrangian coordinates to the Eulerian ones.

Difference models were discussed for different cases of $n$ and
$\gamma$. It is shown that the Samarskii-Popov scheme is  invariant
for the symmetries of the general case of $\gamma$, but  not for the
additional symmetry of the special case $\gamma_* = { n+3 \over
n+1}$. This scheme possesses conservation of mass and energy, for
$n=0$ also conservation of momentum and motion of the center of
mass. It does not have  conservation of the entropy along the
{path}lines. For the special values $\gamma_*$ we suggest
invariant schemes, which have  conservation of mass and conservation of
the entropy along the {path}lines.

\section*{Acknowledgements}

The research was supported by Russian Science Foundation Grant no. 18-11-00238
"Hydrodynamics-type equations: symmetries, conservation laws, invariant difference
schemes".

\end{document}